\documentclass[a4paper,10pt]{article}
\pdfoutput=1
\usepackage[utf8]{inputenc}
\usepackage{jheppubMod}
\usepackage{extarrows}
\usepackage{amsmath,amsfonts,amssymb,graphicx,MnSymbol,subfigure}
\usepackage{hyperref} 
\usepackage{tikz}
\usetikzlibrary{shapes.geometric, arrows}
\usepackage[toc,page]{appendix}

\title{On the Simulation of Hidden Parton Showers in the Conformal Window}
\abstract{We consider confining Hidden Valley/Dark Sector theories containing many dark quark flavors. These theories are in the ``conformal window'': they  reach an infrared fixed point when their quarks are massless, and have unfamiliar confinement when the quark masses are non-zero but small.  Their jets of hidden hadrons may be quite different from those familiar from QCD, but their details cannot currently be simulated even qualitatively. This is partly due to the use of approximations to the two-loop running coupling in existing event generators' parton showers, which are not broadly applicable across the conformal window.  We argue that the exact two-loop running coupling, and a corresponding Sudakov factor employing that coupling, must be implemented in simulation packages in order to allow phenomenological studies of these theories.}

\author[a]{Suchita Kulkarni,}
\author[a]{Joshua Lockyer,}
\author[b]{Matthew J. Strassler}
\affiliation[a]{Institute of Physics, NAWI Graz, University of Graz, Universit\"atsplatz 5, A-8010 Graz, Austria}
\affiliation[b]{Department of Physics, Harvard University, Cambridge, MA, 02138}
\emailAdd{suchita.kulkarni@uni-graz.at}
\emailAdd{joshua.lockyer@uni-graz.at}
\emailAdd{strassler@g.harvard.edu}
\date{December 2024}

\newcommand{\newc}{\newcommand}
\newc{\nc}{{N_C}}                
\newc{\nf}{{N_F}}                
\newc{\ad}{\alpha}                
\newc{\adir}{\ad_*}         
\newc{\aduv}{\ad_{\UV}}         
\newc{\gmIR}{\gamma_{G}}         
\newc{\gmUV}{\gamma_{\UV}}         
\newc{\as}{{\alpha_s}}                
\newc{\ml}{{\mu/\Lambda}}                
\newc{\fml}{\frac{\mu}{\Lambda}}                
\newc{\fc}{{\nf/\nc}}                
\newc{\mzp}{M_{Z^\prime}}                
\newc{\constad}{\xlongequal{\ad = \rm{const}}}%
\newc{\oneloopad}{\xlongequal{\rm{one-loop}\, \ad}}%

\newc{\subt}[1]{\st{#1\ }}
\newc{\UV}{{\mathchoice{}{}{\scriptscriptstyle}{}UV}}
\newc{\IR}{{\mathchoice{}{}{\scriptscriptstyle}{} IR}}
\newc{\mpx}{\sigma}
\begin{document}
\maketitle

\newc{\dela}{{\tilde{\Delta}_{a}}}                
\newc{\epa}{{{\epsilon}_{a}}}
\newc{\delaepa}{{\tilde{\bf \kappa}}}
\newc{\dpe}{{\tilde{\Delta}_{a}}^{{\epsilon}_{a}}} 

\section{Introduction}
\label{sec:introduction}
Both observational data (the apparent existence of dark matter) and theoretical considerations (such as string theory constructions of Standard Model-like vacua, models of broken supersymmetry and models related to the baryon asymmetry) motivate the consideration of interacting hidden sectors: sets of particles that couple to one another but are neutral under the Standard Model (SM) gauge groups.  This very neutrality under the familiar gauge forces makes these particles unobservable in ordinary collider experiments, and it was long assumed that such sectors would mainly be sources of missing energy.  

However, it was pointed out in~\cite{Strassler:2006im, Strassler:2006qa} that if the hidden sector has a mass gap, then this may not be so.  Even if all interactions with the SM are very weak, some of the sector's particles may decay back to observable SM particles with lifetimes that make their decays visible in experiments.  Such interactions can occur through neutral ``portals'', which might include couplings between SM and hidden Higgs bosons,  mixing of SM and hidden (i.e. sterile) neutrinos, a neutral loop of particles charged under both sectors (such as quirks), or mixing of a spin-one hidden particle with the photon, the $Z$, or a new neutral vector boson from beyond the Standard Model (BSM).  Such experimentally-visible hidden sectors are most often referred to as ``hidden valleys'' (HV) and/or interacting ``dark sectors'' (DS); we will treat these terms interchangeably and will abbreviate them as HV/DS. 

Theoretical HV/DS models have a long history --- this includes the Twin Higgs and its variants~\cite{Chacko:2005pe,Craig:2015pha, Craig:2014aea, Craig:2016kue, Craig:2014roa} --- and more continue to be invented. 
But from a general and purely experimental perspective, HV/DS models are interesting and challenging because, as observed  in~\cite{Strassler:2006im}, they produce many experimental signatures not seen in the SM.  This is especially true for confining HV/DS models, where physics involving both resummed perturbation theory and fully non-perturbative effects can lead to high multiplicity final states, unusual clustering of particles, and/or long-lived particles.  A number of these novel signatures, some of which have recently been given names such as semi-visible jets~\cite{Cohen:2015toa}, trackless jets~\cite{bai2011dark}, emerging jets~\cite{Schwaller:2015gea} and soft-unclustered energy patterns~\cite{Strassler:2008bv, Hofman:2008ar, Hatta:2008tx, Knapen:2016hky, Harnik:2008ax} in theoretical studies, have been sought by the ATLAS, CMS and LHCb experiments~\cite{CMS:2021dzg, CMS:2024gxp, ATLAS:2023swa, CMS:2024nca, ATLAS:2023kao, CMS:2018bvr} at the Large Hadron Collider (LHC).  
Furthermore, HV/DS bound states can also be an attractive dark matter [DM] candidate~\cite{Hochberg:2014kqa, Hochberg:2015vrg,Kulkarni:2022bvh, Pomper:2024otb, Bernreuther:2023kcg}.  For reviews on HV/DS theories with DM candidates, see e.g.~\cite{Albouy:2022cin,Kribs:2016cew,Cacciapaglia:2020kgq}. 

However, most of the LHC and DM studies either have been targeted at toy ``simplified models'' or have focused on confining HV/DS models whose physics all resemble that of real-world QCD.  More precisely, the confining hidden sectors most often considered exhibit a ``dark'' parton shower, ``dark'' hadronization, and a ``dark hadron'' spectrum that are assumed to resemble the showering, hadronization and spectrum that we are familiar with in QCD.  It makes sense in initial studies to restrict attention to QCD-like theories, since the farther that confining sectors stray from QCD-like behavior, the less we understand them and the more limited our ability to simulate them.  But we must find ways to move beyond this restriction, since nature has no reason to respect it.

In this paper, we consider the challenges of dark showers that are qualitatively different from QCD showering because of the unfamiliar running of the HV/DS gauge coupling.  We will focus on non-Abelian gauge theories that resemble QCD in having an $SU(\nc)$ gauge group with $\nf$ flavors of quarks and antiquarks in the fundamental and anti-fundamental representation, but where $\fc$ is substantially larger than in our own QCD sector.  Using well-known facts about the two-loop beta function in such theories, we will show at a theoretical level that existing simulation tools are not currently able to handle this case, and give a discussion of what would be needed to improve them.  A companion paper in preparation will discuss the practical aspects of implementing these improvements into the Hidden Valley module~\cite{Carloni:2011kk,Carloni:2010tw} of PYTHIA 8~\cite{Sjostrand:2007gs, Bierlich:2022pfr}.  (See also recent work on a Hidden Valley module for Herwig \cite{Kulkarni:2024okx}.)

\subsection{Phenomenological Motivation}
Much of our attention will be focused on the ``conformal window'' (CW), the region of $\fc$ where, if all hidden quark flavors are massless, the theory flows to an infrared fixed point (IRFP). This is not to say that these theories are automatically conformal --- at a generic value of the coupling constant $\ad$, the coupling does run --- but it does mean that if one were to set the UV value of the coupling constant to be equal to its fixed-point value $\ad_*$, then the resulting theory would be strictly conformal. 

Our phenomenological interest is not in conformal theories, or even in a running coupling with massless quarks, but instead in the situation where some or all quark masses $M_q$ are non-zero.   In many such theories, even though the ultraviolet (UV) behavior is that of a theory in the CW window and may approach an IRFP, the deep infrared (IR) behavior at scales below $M_q$ is that of a confining theory.  In such a case, HV/DS phenomenology will be generated in the IR, but its shape and details will be affected by the unfamiliar, non-QCD-like behavior in the UV.

\begin{figure}[h!]
\centering 
    \includegraphics[width=0.45\textwidth]{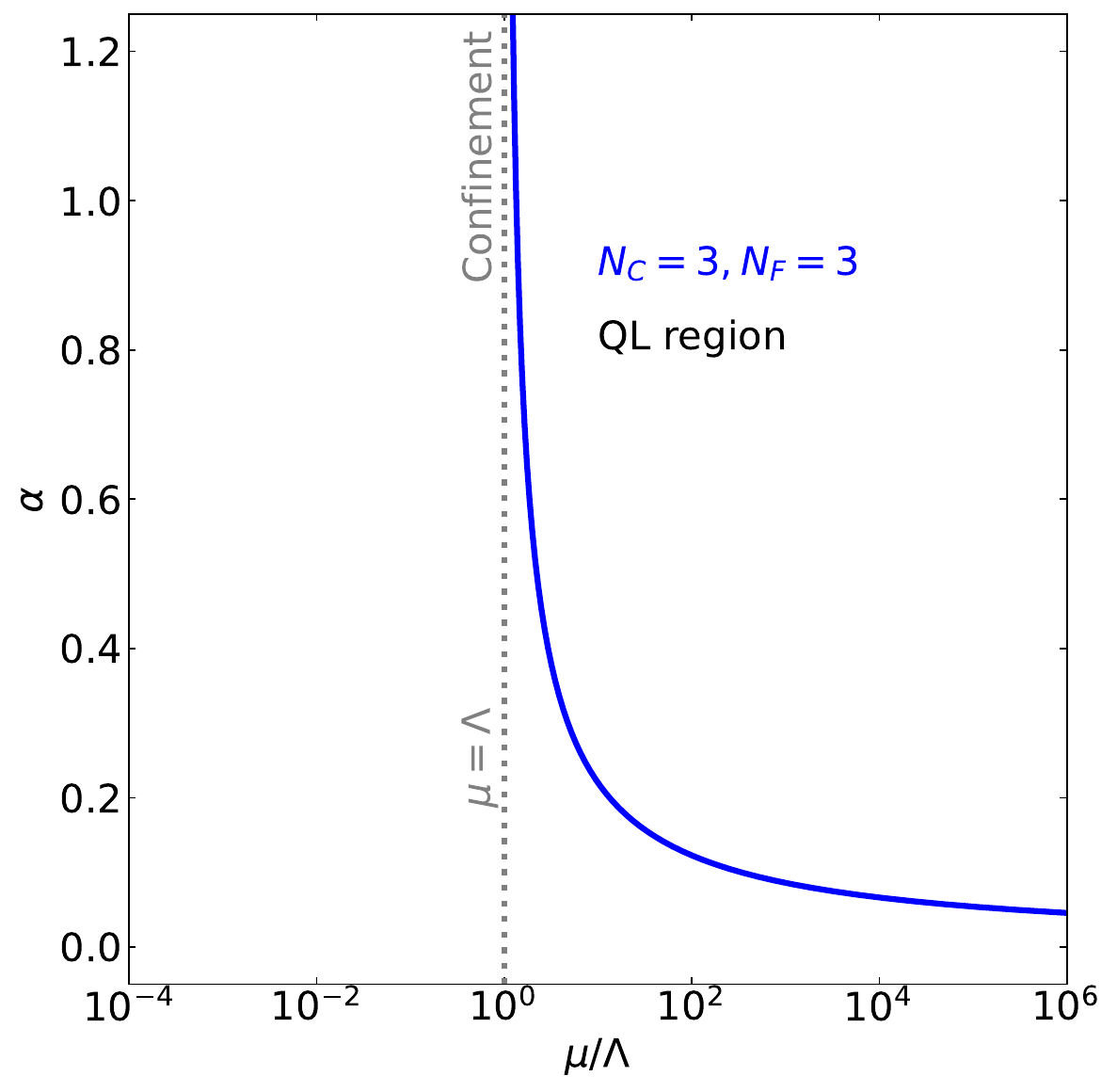}    \includegraphics[width=0.45\textwidth]{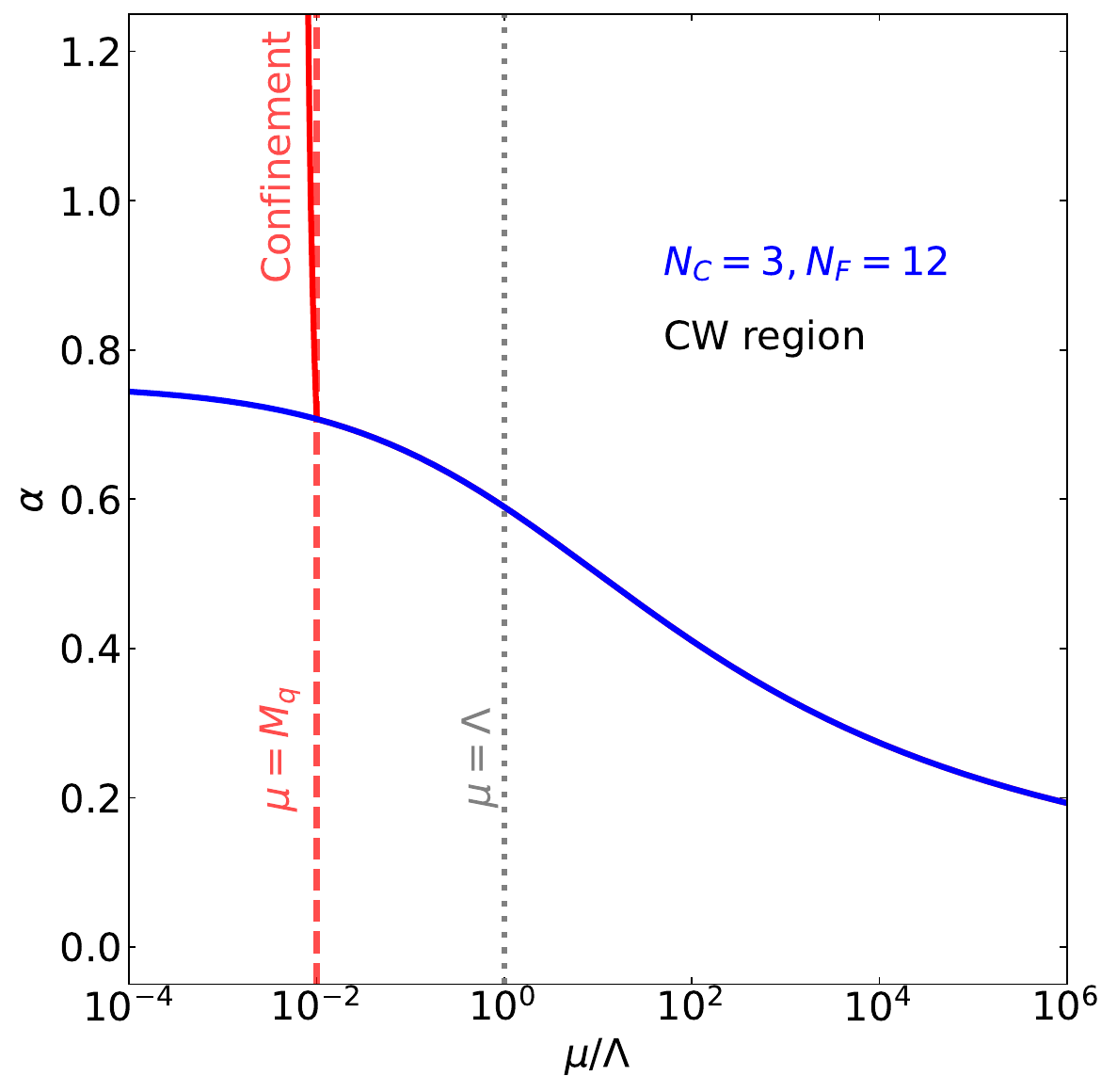}
    \caption{(Left panel) In the QCD-like region, the two-loop running coupling runs steadily to infinity. We take $\nc=3, \nf=3$ as an example.  (Right panel) In the conformal window region, for quark mass $M_q=0$, the two-loop coupling approaches an infrared fixed point (blue curve), while for  $0< M_q\ll\Lambda$ the quarks decouple and the coupling diverges (red curve) just below $\mu= M_q$. We take $\nc=3, \nf=12$ as an example.  Confinement and hadron formation are expected to occur at roughly the scale where the two-loop coupling diverges.}
    \label{fig:sketch_alpha}
\end{figure}

In the left panel of fig.~\ref{fig:sketch_alpha}, the familiar running of the coupling constant $\ad$ in a QCD-like theory is shown; the coupling blows up at the scale $\Lambda$, which is roughly where confinement sets in.  The right panel displays the running of $\ad$ for a theory in the CW. For $M_q=0$ the coupling approaches a constant in the IR.  But if all or many of the quarks have mass $M_q>0$, enough to push the IR theory out of the CW, then the coupling blows up just below $M_q$. Confinement then occurs around this scale. 

Notice that, in this case, $\Lambda$ is not the scale of confinement.  Instead it characterizes the crossover from the familiar one-loop logarithmic running coupling to the approximate fixed point regime.  
The confinement scale is instead set by a non-trivial combination of $M_q$ and $\Lambda$, and lies just below $M_q$ if $M_q\ll\Lambda$.  

Comparing the left and right panel, we see an important effect: when $\mu$ is somewhat above the confinement scale, the value of the coupling constant can be much larger in a CW theory with $M_q>0$ than in a theory with QCD-like confinement.  Consequently the parton shower, even well above the confinement scale, may potentially be quite different in the two regimes. 
This will lead to quantitative differences, and perhaps even qualitative ones, between the phenomenological signatures of QCD-like HV/DS models and those in the CW window.

To ensure complete experimental coverage of these types of theories, it is essential to check whether existing searches are sensitive to these less familiar HV/DS signatures and, if not, to extend existing search strategies or invent new ones.   Doing so requires simulation tools for such models, motivating the present study.
 
\subsection{Classification of Models}
\label{subsec:model_classification}
We next give a rough classification of the theories in question.  As is well known, their qualitative features depend mainly on $\fc$ if $\nc\gg 1$.  The features at large $\nc$ are expected already to be largely true for $\nc=3$, a point on which many simulation tools rely.  We therefore discuss this classification in the large-$\nc$ regime, implicitly assuming that it applies also for $\nc=3$.

Such theories have a positive one-loop beta function, and are thus IR free (IF), when $\fc \geq 5.5 \equiv (\fc)_{IF}$.  For $\fc$ just below $(\fc)_{IF}$, where the one-loop beta function is negative but small, the existence of IRFPs can be established using two-loop perturbation theory,  because the two-loop beta function has a zero  which higher-loop corrections cannot remove \cite{Caswell:1974gg}.  Such fixed points are often called ``Banks-Zaks'' fixed points following \cite{Banks:1981nn}.  Thus the existence of a CW region has long been established.  

We will refer to the lower limit of the CW region as $(\fc)_{CW}$. Neither general theoretical arguments nor lattice gauge theory (LGT) simulations can currently establish the numerical value of $(\fc)_{CW}$ or determine what happens just below it. In ${\cal N}=1$ supersymmetric QCD, the CW region was discovered decades ago to be much larger than the BZ region \cite{Seiberg:1994pq}. 
Evidence that the CW region in non-supersymmetric QCD extends far beyond the BZ region, down to $(\fc)\sim 2 - 3$, has been given in simulations using LGT as well as other non-perturbative approaches.

For an early review on these LGT efforts, see \cite{Fleming:2008gy} with~\cite{Drach:2020qpj} providing more recent updates. Numerous LGT studies aim to pinpoint the lower end of CW region~\cite{LatticeStrongDynamics:2023bqp, LatKMI:2013bhp, Cheng:2013xha, Aoki:2012eq, LSD:2023uzj, Hasenfratz:2023wbr, Hasenfratz:2024fad, Hasenfratz:2017qyr, Fodor:2019ypi, Kuti:2022ldb}. These studies have mainly focused on the $SU(3)$ gauge group, and suggest that the lower end of conformal widow is somewhere between 8 and 10 flavors. This expectation is in accordance with results~\cite{Gies:2005as, Chung:2023mgr} using other non-perturbative methods.\footnote{Studies using functional methods alone show the lower end of conformal window at  $\nf \approx 4.5$~\cite{Hopfer:2014zna}, which is at odds with other results. }

The actual value of $(\fc)_{CW}$ will not be central to our discussion.  When necessary we will use the two-loop value, which is $(\fc)_{CW}=2.62$ at $\nc \to \infty$, knowing that the true value will be somewhat different.  Also, though  it will not be essential in our discussion, we will assume for simplicity that below the CW region all theories are QCD-like, in the sense that they confine and (for $\nf>1$) exhibit chiral symmetry breaking in much the way real-world QCD does. There may be other regimes that lie between these two, but this will not affect our main points.

\begin{figure}[h!]
\centering 
    \includegraphics[width=0.9\textwidth]{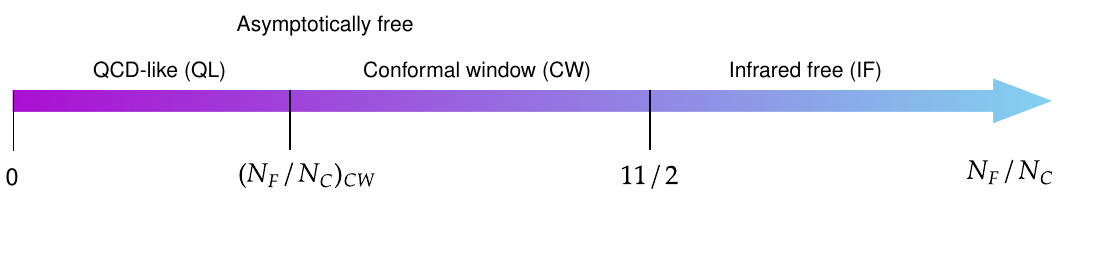}
    \caption{Characteristic behavior of theories of $\nc$ colors and $\nf$ quarks as a function of $\fc$, under the assumptions made in this paper that the conformal window and QCD-like regime meet without any additional regimes. The region just below $\fc = 11/2$ is known as the Banks-Zaks region, where the magnitude of the fixed point coupling becomes small. The value of $(\nf/\nc)_{CW}$ is currently unknown. }
    \label{fig:phase_diag}
\end{figure}

In short, for the limited purposes of this paper, we assume that these theories with $\nc$ colors and $\nf$ flavors exhibit three main behaviors, as sketched in fig.~\ref{fig:phase_diag}:
\begin{itemize}
	\item In the infrared-free (IF) region, with  $\fc > (\fc)_{IF} = 5.5$, the theories are infrared free and their IR physics can be treated using one-loop perturbation theory alone.
	\item In the QCD-like (QL) region, with $\fc < (\fc)_{CW}$, the theories exhibit QCD-like showering and QCD-like confining behavior in the infrared; the full range of methods used in simulating QCD are needed to study them. 
	\item In between we find the conformal window (CW) region, where there is an infrared fixed point for $M_q=0$ and an unfamiliar pattern of showering.
\end{itemize}

\subsection{Value and Limitations of a Two-Loop Study}
The long-term goal of our investigations is to allow for the exploration of phenomena found in the conformal window but absent in QCD-like theories. 
At a minimum, any such study must be able to capture the qualitative features that are expected to be present in the full theory, as seen in the right panel of fig.~\ref{fig:sketch_alpha}: 
\begin{itemize}
	\item that $\alpha$ grows logarithmically from a small value, as determined at one loop;
	\item that after $\alpha$ grows sufficiently, its running slows due to higher order effects;
	\item that $\alpha$ approaches an infrared fixed point value $\alpha_*$, at a rate controlled by an anomalous dimension $\gamma$.
\end{itemize}
The last two features are not visible at leading order (LO), but do appear at next-to-leading order (NLO).  

Any method of event generation which captures these features will have to evaluate the running coupling at two loops or higher. While a fully consistent higher-order parton shower still lies in the future,  the most naive option, which is to combine a leading-order parton shower with the two-loop running coupling, may already be enough to identify potential weaknesses in experimental search techniques.  

Not even this option can be carried out today, however, due to a minor technical obstruction. Currently, event generators that evaluate the two-loop running coupling use  approximation schemes that work in the QL regime but not across the CW regime, even for $\mu$ well above $\Lambda$. This prevents study of the most interesting phenomena that arise in fig.~\ref{fig:sketch_alpha}.

 For example,  PYTHIA 8 uses an approximation (the so-called ``PDG formula") to the two-loop running coupling, rather than its known exact form.  Derived as an ultraviolet expansion, this formula works well for real-world QCD with $\fc\sim 1$. But in the CW regime, it fails for two reasons: first, it is only potentially valid in the ultraviolet, and so cannot capture either the IRFP or the crossover region, and second, more surprisingly, it often fails to be accurate even for $\mu$ well above $\Lambda$.   We will see this explicitly below.

To evade this problem is straightforward in principle: PYTHIA and other generators should use the full two-loop running coupling. At first glance this may seem a trivial point, as the solution to the two-loop beta function is well known in closed form~\cite{Gardi:1998qr, Gardi:1998rf, Magradze:1999um,Prosperi:2006hx}.  The two-loop $\alpha(\mu)$ can be expressed in terms of the Lambert function $W(z)$, where $z$ is a single variable that depends on $\fc$ and $\mu/\Lambda$. The function has two real branches, $W_{-1}(z)$ and $W_0(z)$, whose relevance depends on $\nf/\nc$.

However, any event generator must compute the coupling efficiently, since it must be evaluated multiple times in each parton shower.  Optimized code for the Lambert function is not readily available --- it does not appear in standard math packages --- so we must address the practical question of how best to compute it. A lookup table would be unwieldy, as the range of $z$ over which the function must be evaluated is enormous. Meanwhile, simple infrared and ultraviolet expansions of $W(z)$ do not overlap. Numerical calculation of the Lambert function has been considered in~\cite{Veberic:2010ay,Veberic:2012ax}, with emphasis on mathematical precision.\footnote{Approximations are given that, taken together, have accuracy for several decimal places for both $W_{0}(z)$ and $W_{-1}(z)$.  The Vincia shower in PYTHIA uses one of these approximations, although it applies only for a limited range of $z$.} Our focus here will be on physically-motivated expansions of this function, which clarify its physical application in the CW regime, and we will postpone the practical issues of implementation to future work. 

One other important issue is that parton showers require computation of Sudakov factors.  Again there is a technical obstruction: certain event generators, most notably PYTHIA, evaluate the Sudakov factor using veto algorithms whose assumptions are valid in the QL regime but not in the CW regime.  We will show how this issue can be resolved, providing useful formulas that can be used in any veto algorithm.

To reiterate, neither one-loop nor currently-implemented approximate two-loop running couplings can be used to study HV/DS models in the CW regime, even at a qualitative level.  Our immediate goal is to rectify the situation. Of course, any two-loop approximation will itself be subject to higher-loop ~\cite{Ryttov:2010iz,Ryttov:2012nt,Ryttov:2013ura,Ryttov:2014nda,Gracey:2023unc} and non-perturbative corrections, which are  scheme-dependent and often large outside the BZ region.  Nevertheless, detailed study of CW models cannot begin until this first step is complete. 

After some preliminary discussion of the QL and CW regions in section~\ref{sec:RGE_introduction}, we will discuss the exact two-loop coupling, and the pros and cons of various approximations to it, in  section~\ref{sec:explicit_solutions}.  We address the computation of the Sudakov factor in section~\ref{sec:two_loop_sudakov}, supplemented by appendix~\ref{app:sudakov}. We conclude with a brief discussion, including the potential impact of higher-order effects, in section~\ref{sec:conclusion}.

\section{Properties of the renormalization group equation (RGE)}
\label{sec:RGE_introduction}
A HV/DS with an $SU(\nc)$ gauge group and $\nf$ Dirac fermions in the fundamental representation has Lagrangian 
\begin{equation}
	\mathcal{L}_\text{UV} = -\frac{1}{4}G_{\mu\nu}^aG^{\mu\nu, a} + \bar{q}(i\gamma^\mu D_\mu - M_{q})q \, ,
\end{equation}
where $M_{q}$ is the mass of the dark quarks, $G^{\mu\nu}$ denotes the dark gluon field strength tensor and $D_\mu$ is the gauge covariant derivative. $G^{\mu\nu}$ is given by,
\begin{equation}
    G_{a}^{\mu\nu} = \partial^\mu G_a^\nu -\partial^\nu G_a^\mu -g f^{abc} G_b^\mu G_c^\nu,
\end{equation}
where $a,b,c$ denote color indices, $f^{abc}$ denote the totally anti-symmetric structure constants and $g$ is the energy dependent gauge coupling. We define the running coupling as $\ad = g^2/(4\pi)$ throughout our investigation. Although $\ad$ is a function of the energy scale $\mu$ and of $\fc$, we will rarely notate this dependence explicitly. 

In the limit $M_{q} \ll \Lambda$, the RGE for $\ad$ can be written 
\begin{equation}
    \mu^2\frac{d\ad}{d\mu^2}= \beta\left(\ad\right) = -\ad^2 \displaystyle\sum_{n=0} ^{\infty}\beta_n \ad^n ,
    \label{eq:RGE}
\end{equation}
where $\beta_n$ are the $(n +1)$-loop beta function coefficients.\footnote{We follow the convention where the $\beta_n$ coefficients have explicit factors of $4\pi$. This differs from the convention used in \cite{PDBook}, where these factors are absorbed into the $\beta$ function definition.} 
This equation can be solved with a boundary condition $\ad=\ad_0$ at a reference scale $\mu=\mu_0$. 

The first two $\beta_n$ coefficients are  
\begin{eqnarray}
    \beta_{0} &=& \frac{1}{4\pi}\left(\frac{11}{3} C_A - \frac{4}{3} T_R \nf\right) ,\nonumber \\
    \beta_{1} &=& \frac{1}{\left(4\pi\right)^2}\left(\frac{34}{3}C_A^2 - 4 C_F T_R \nf -\frac{20}{3} C_A T_R \nf\right).
    \label{eq:betafunctions} 
\end{eqnarray}
Here $C_A = \nc$ and $C_F = {\left(\nc^2 - 1\right)}/{\left(2 \nc\right)}$ are the adjoint and fundamental Casimir invariants, while $T_R = 1/2$. (The $\beta_n$ coefficients for $n > 2$ are scheme dependent.) In general, $\beta_n = \nc^{n+1} f(\fc)[1 + {\rm order}(1/\nc)]$, and there is a large-$\nc$ expansion in which one takes $\nc\to \infty$ holding the `t Hooft coupling  $\ad(\mu)\nc=h(\fc,\ml)[1+\mathcal{O}(1/\nc)]$ fixed. We will usually show results at large $\nc$ for simplicity, where $\nc$ scales out and physics (including $\ad(\mu)\nc$) depends only on $\fc$ and $\mu/\Lambda$.
 
If there exists a zero of the full beta function at $\ad=\ad_*$ ({\it i.e.}, if $\beta\left(\ad_*\right) = 0$ non-perturbatively), then there exists a conformally invariant theory with $\ad=\ad_*$ at all scales.  If $\beta(\ad)<0$ for $\ad<\ad_*$, this fixed point is an attractive IRFP, in that $\ad$ will run to $\ad_*$ in the IR.  At two loops, this is the situation throughout the CW region. 

However, at any fixed order, the presence and location of zeros of the beta function are order- and scheme-dependent. All results below  will be computed only at two loops. Whether there does or does not exist a physical, non-perturbative fixed point for a particular choice of $(\fc)$ remains  an important open question.  

At one-loop, eqn.~\eqref{eq:RGE} can be solved exactly and the resulting $\ad$ diverges at a finite scale $\Lambda = \mu_0\exp(-\beta_0{\ad_0}/2)$, signaling a breakdown of perturbation theory. The one-loop coefficient $\beta_0$ famously switches sign at $\fc = (\fc)_{IF} = 5.5$ and so only cases with $\fc<5.5$ are  asymptotically free.

The RGE has characteristically different behaviour at two loops if $\fc$ is large enough. For small $\fc$, both $\beta_0$ and $\beta_1$ are positive and the beta function is negative everywhere; this is the QL region where the coupling runs analogously to its behavior in QCD.  But $\beta_1$ changes sign at \cite{Caswell:1974gg, Jones:1974mm}
\begin{equation}
    \displaystyle\left(\frac{\nf}{\nc}\right)_{CW} = \displaystyle\frac{34}{\left(13-\frac{3}{\nc^2}\right)} \  \to \ 2.62  \ \ \ (\nc\to\infty).
    \label{eq:IRFPonset}
\end{equation}
 (For $\nc=3$,   $(\fc)_{CW}=2.68$.)  
Here we enter the (two-loop) CW region, since for larger $\fc$, the two-loop $\beta$ function has an IRFP at  \cite{Caswell:1974gg,Banks:1981nn}
\begin{equation}
	\ad_{*} = -\displaystyle\frac{\beta_0}{\beta_1}.
\end{equation}
As the theory approaches an IRFP, $\ad$ approaches $\ad_*$ as a power of $\mu$.  This is as it must be, since gauge invariant operators in a conformal theory have definite scaling dimensions, and the dimension of the least irrelevant operator controls the flow into the IRFP.  We define this critical exponent as \cite{Peterman:1978tb, balian1981methods} 
\begin{equation}
    \gamma = \displaystyle\frac{\partial\beta}{\partial\ad}\bigg\rvert_{\ad = \ad_*},
    \label{eq:criticalexponent}
\end{equation}
which at two loop is
\begin{equation}
    \gamma = \displaystyle\frac{-\beta_0^2}{\beta_1} = \beta_0\ad_*.
    \label{eq:gamma_twoloop}
\end{equation}
Our definition implies that $\gamma$ is half the anomalous dimension of the leading irrelevant operator at the fixed point (which includes the square of the gluon field strength, Tr($G_{\mu\nu}G^{\mu\nu}$), plus other terms), and therefore
\begin{equation}
    \ad_*-\ad \propto  \mu^{2\gamma} 
\end{equation}
as $\mu\to 0$.

\begin{figure}[h!]
\centering 
    \includegraphics[width=0.45\textwidth]{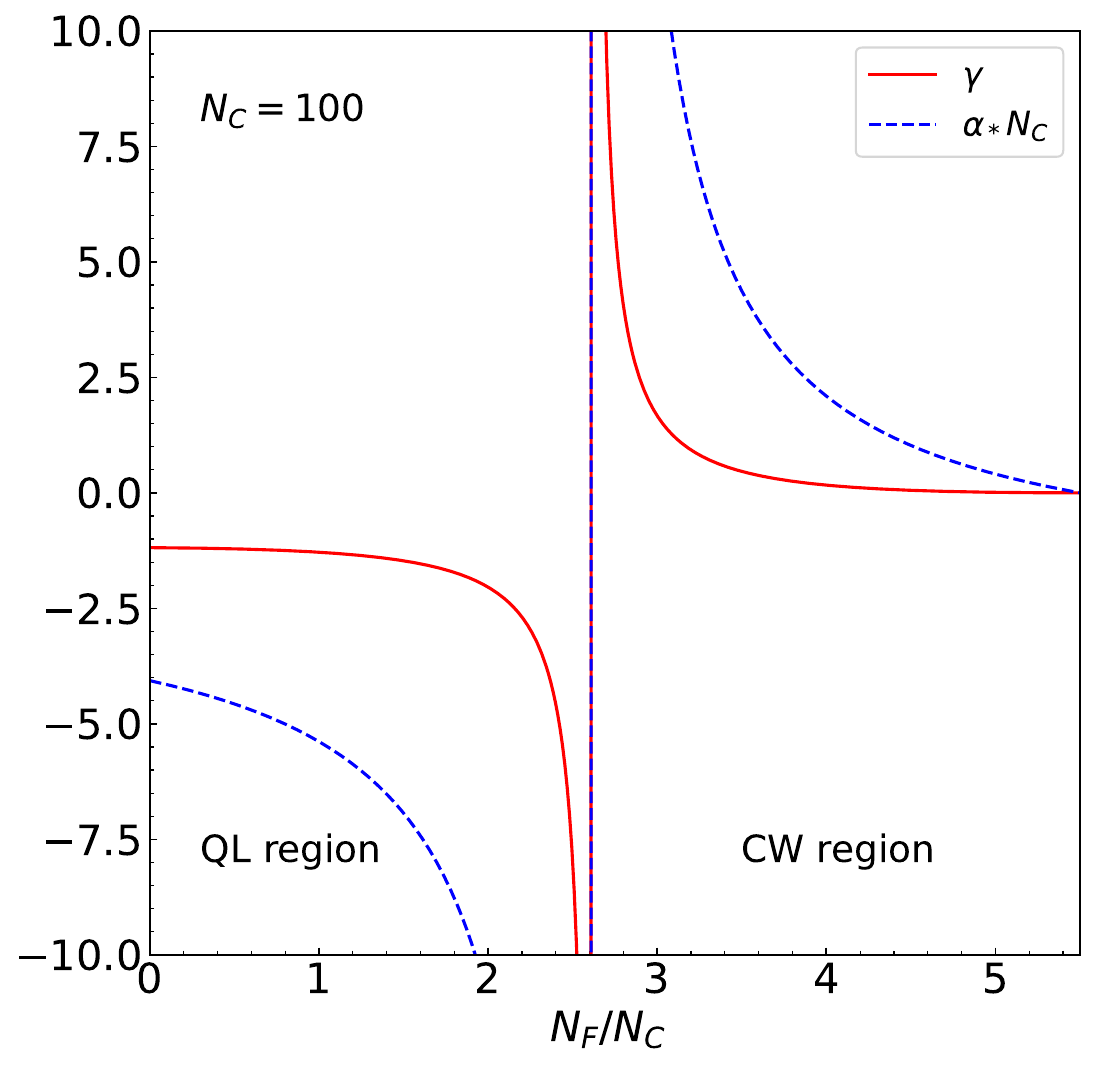}
    \caption{Characteristic two-loop behavior of the fixed point 't Hooft coupling $\ad_* \nc$ and the anomalous dimension $\gamma$ as a function of $\fc$ in the CW.  Their analytic extensions into the QL region are also shown.} 
    \label{fig:gamma_alphastar}
\end{figure}

Our definitions of $\ad_*$ and $\gamma$ remain useful even when continued into the QL region, where $\ad_*<0$, $\gamma<0$, and there is no physical fixed point.  The dependence of these two quantities on $\fc$ is shown in fig.~\ref{fig:gamma_alphastar}; we plot the fixed points' `t Hooft coupling $\ad_* \nc$ rather than the gauge coupling because the former is $\nc$-independent at large $\nc$. Note that both $\ad_* \nc$ and $\gamma$ go to zero in the BZ region, but not elsewhere; this is the only regime in which two-loop calculations will not receive substantial corrections.  At this order, both quantities diverge at $(\fc)_{CW}$.

Note that the coupling constant is always scheme-dependent, but $\gamma$, related to the dimension of an observable operator in a fixed point theory, is physical and thus scheme-independent.   For this reason it may often be useful to specify the CW fixed points by their value of $\gamma$, rather than their value of $\ad_*$, since the former is physical and could be calculated non-perturbatively, perhaps using LGT.  That said, any calculation of $\gamma$ as an expansion in $\ad$, such as we are able to do here, inherits the coupling's scheme-dependence at higher orders.  In particular, one cannot trust the location (and even the existence) of the divergence of $\gamma$ in fig.~\ref{fig:gamma_alphastar}.

\section{Explicit solutions to the RGE}
\label{sec:explicit_solutions}

In this section we review the known exact form of the two-loop running coupling, and consider various approximations to it.  Our focus will be on the usefulness and limitations of these approximations. 

\subsection{Review of exact solutions of the two-loop RGE equation}
\label{sec:exact_solutions}

The two-loop running of $\ad$ is obtained by integrating eqn.~\eqref{eq:RGE} truncated to second order.  The exact solution can be written~\cite{Prosperi:2006hx, Gardi:1998qr, Gardi:1998rf, Magradze:1999um} in terms of the Lambert $W$ function
\begin{equation}
    \ad(\mu) = \displaystyle\frac{\ad_*}{1 + W(t)}\ ,
    \label{eq:runnWfn}
\end{equation}
where 
\begin{equation}
	t = \left(\frac{\ad_*}{\ad_0} - 1\right)e^{\ad_*/\ad_0 - 1}\left(\frac{\mu}{\mu_0}\right)^{2\gamma} \ . 
	\label{eq:tdef1}
\end{equation} 
Here $\alpha_0\equiv \alpha(\mu_0)$ with $\mu_0$ a reference scale.\footnote{To see that eqns.~\eqref{eq:runnWfn} and \eqref{eq:tdef1} are consistent when $\mu=\mu_0$ requires the Lambert function identity $x= W(x)\exp{[W(x)]}$~\cite{Corless:1996}. }
This form is valid in both the QL and CW regions, though to obtain real and positive-valued solutions that asymptote to one-loop running in the UV, the correct branch of the Lambert  function must be chosen for each region. For real $t$, the Lambert $W$ function has two real branches, $W_0$ (the principal branch) for $t \geq 0$, which is appropriate in the CW region, and $W_{-1}$ for $-1/e \leq t < 0$, relevant for the QL region.

As is commonly done in the Standard Model, one can use eqn.~\eqref{eq:runnWfn} to uniquely define the running coupling given $\ad_0, \mu_0$. In that context, $\nf$ and $\nc$ are fixed, and $M_Z$ provides a known and fixed non-QCD scale at which $\alpha_s(M_Z)$ can be defined.  But for the purposes of studying a HV/DS, this approach is generally inconvenient.  This is because the experimental task is to search across many possible HV/DS sectors, and involves looking for signatures of new dark sector particles with definite masses. While the scale of the typical dark hadron masses is exponentially sensitive both to $\alpha(\mu_0)$ and $\fc$, it is only power-law sensitive to the infrared scales, which may include the dimensional-transmutation scale $\Lambda$ and low quark masses, as  fig.~\ref{fig:sketch_alpha} illustrates.\footnote{In addition, not only is there is no natural model-independent choice of $\mu_0$, any additional mass thresholds near the scale $\mu_0$ would affect $\alpha(\mu_0)$ while having no observable effect on experiments.}  Experimental results will therefore best be characterized by the use of these scales rather than $\alpha(\mu_0)$. For this reason we now exchange $\alpha_0$ and $\mu_0$ for $\Lambda$.

The procedure for specifying $\Lambda$ is familiar in the QL region \cite{Prosperi:2006hx,PDBook} where $\Lambda$ may be defined to be the scale at which $\ad$ diverges.  This is also approximately the scale at which confinement occurs. Specifically, by setting $t = -1/e$ and $\mu = \Lambda$, we get
\begin{equation}
    \beta_0\ln\left(\frac{\Lambda^2}{\mu_0^2}\right)= -\frac{1}{\ad_0} - \frac{1}{\ad_*}\ln\left(1-\frac{\ad_*}{\ad_0}\right)\ ,
    \label{eq:lambdaqcdlike}
\end{equation}
whose exponentiated form is 

\begin{equation}
    \left(\frac{\ad_*}{\ad_0} - 1\right)e^{\ad_*/\ad_0 }\left(\frac{\Lambda}{\mu_0}\right)^{2\gamma} = -1 \ .    
    \label{eq:QCDLambda}
\end{equation}
With this definition, $t$ simplifies and can be written as 
\begin{equation}
    -t = z = \frac{1}{e}\left(\frac{\mu}{\Lambda}\right)^{2\gamma} > 0\ .
    \label{eq:argumentQL}
\end{equation}
Recall that both $\ad_*$ and $\gamma$ are negative in the QL regime. Consequently $z \ll 1/e$ corresponds to the UV, while the divergence in the coupling occurs at $z = -t = 1/e$, that is, at $\mu=\Lambda$.

By contrast, in the CW region, where both $\ad_*$ and $\gamma$ are positive, the coupling does not diverge in the IR, and so $\Lambda$ must be defined in another way. If $\ad_0 \ll \ad_*$ in the UV, $\Lambda$ should represent the scale at which the coupling ceases to run logarithmically and approaches the IRFP; see fig.~\ref{fig:sketch_alpha}.
It proves useful to analytically continue the form from the QL regime, taking $t=z$ instead of $-z$ and setting $\Lambda$ to be the value of $\mu$ when $t = z = 1/e$.
This gives us
\begin{equation}
    \left(\frac{\ad_*}{\ad_0} - 1\right)e^{\ad_*/\ad_0}\left(\frac{\Lambda}{\mu_0}\right)^{2\gamma} = 1\     
    \label{eq:IRFPLambda}
\end{equation}
for $\ad_0 < \ad_*$ \cite{Appelquist_1996,Appelquist_1999}.   With this definition of the characteristic scale $\Lambda$, we have
\begin{equation}
    t = z \equiv \frac{1}{e}\left(\frac{\mu}{\Lambda}\right)^{2\gamma} > 0 \ .
    \label{eq:argumentCW}
\end{equation}
Note that $z$ is the same as in the QL regime.  However, here $z\to\infty$ is the UV regime, and $z\to 0$ is where $\mu\to 0$ and $\ad\to\ad_*$. 

In summary, when $\ad$ is zero in the UV, the exact two-loop running coupling is 
\begin{equation}
    \frac{\ad_*}{\ad} = \displaystyle
    \Bigg\{
    \begin{matrix}
    {1 + W_{-1}(-z)}  \ \ \ \  &({\rm QL}) \cr {1 + W_{0}(z)} \ \ \ \ &({\rm CW})
    \end{matrix} \ .
    \label{eq:runnWfn_Lambda}
\end{equation}
We reiterate that in the CW regime,  $\gamma>0$ (and $\ad_*>0$) and the UV is at $z\gg 1/e$, whereas in the QL regime, $\gamma<0$ (and $\ad_*<0$) and the UV is at $z\ll 1/e$.

The treatment of the QL and CW regions can be further aligned if we define
\begin{equation}
    v=\Bigg\{
    \begin{matrix}
    1/z  \ \ \ \  &({\rm QL}) \cr z\ \ \ \ &({\rm CW})
    \end{matrix} \ ,
    \label{eq:LamW_argument}
\end{equation}
so that the UV is at $v\gg 1/e$ in both regimes. In a moment we will clarify the nontrivial relations, as a function of $\fc$, between $v$, $\gamma$, and $\alpha \nc$. 

For completeness, we mention three other regimes where the coupling takes a different form.  In the IF regime ($\fc\geq 5.5$), $\ad_*<0$ but $\gamma>0$, so $z\ll 1/e$ now corresponds to the IR ($\mu\ll \Lambda$). As in the QL region, the coupling diverges at $z=-t=1/e$ ($\mu=\Lambda$) but in this phase it is now a Landau pole in the UV. Additionally, in the IR, as $z\to0$ ($\mu\to0$), the running coupling flows to $\alpha=0$. 

In the CW regime, one may imagine setting $\ad_0=\ad_*$ exactly.   In this case $t=0$ for all $\mu$;  the theory is exactly conformal and $\Lambda$ is not defined.  

Also in the CW regime, one may consider $\ad_0 > \ad_*$ in the ultraviolet.  In this case $\Lambda$ represents a Landau pole, as in the IF region.  Specifically, $t$ is negative, and it is more appropriate to define $z$ to be the same as in the QL region; see eqn.~\eqref{eq:argumentQL}.  This gives the following solution for $\ad$: 
\begin{equation}
    \frac{\ad_*}{\ad} = {1 + W_{0}(-z)} \ .
    \label{eq:running_UVLandau}
\end{equation}
This running coupling still approaches $\ad_*$ in the IR, but diverges in the UV at $\mu=\Lambda$. While this phase of the theory, with some physical UV cutoff, may potentially lead to interesting phenomenology of its own, we will not consider it in this paper. 
\subsection{Relations among important quantities}
\label{sec:relations}

\begin{figure}[h!]
\centering     
    \includegraphics[width=0.4\textwidth]{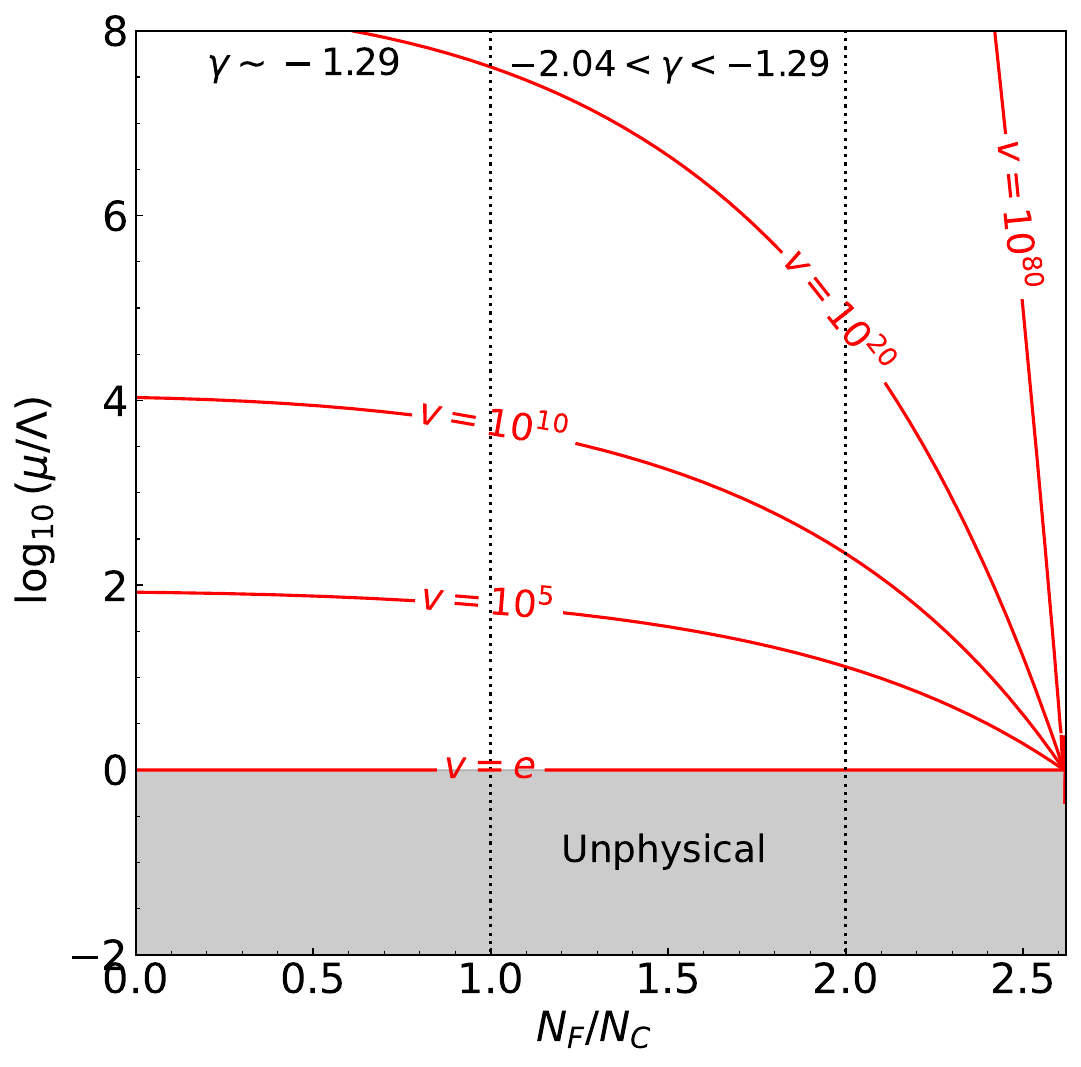}
    \includegraphics[width=0.4\textwidth]{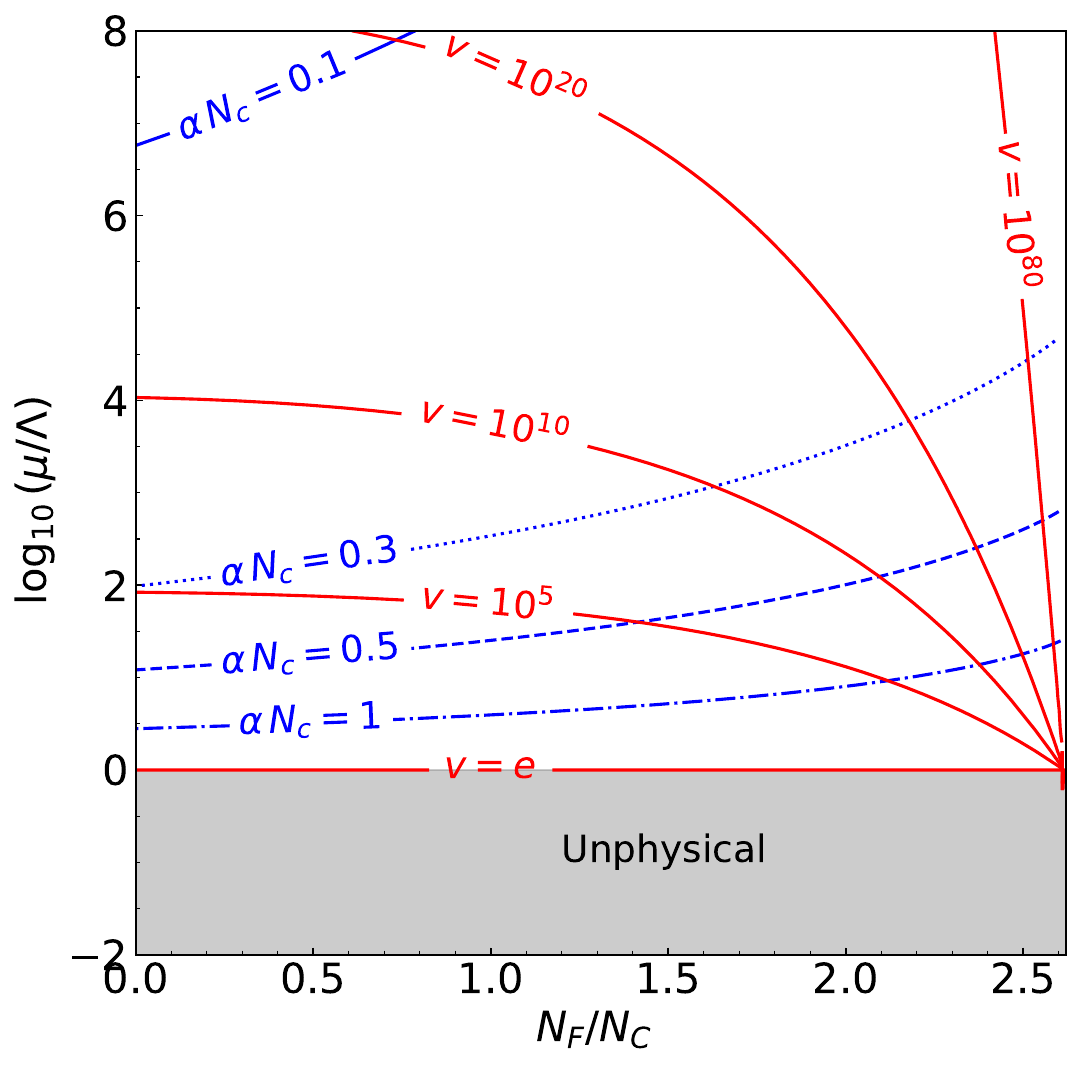}
    \caption{In the QL region, the relationship between  $v$ and $\gamma$ (left panel) and between $v$ and the running 't Hooft coupling $\ad \nc$ (right panel) as a function of $\fc$ and  $\ml$. Note the definition of $v$ in eq.~\eqref{eq:LamW_argument}.}
    \label{fig:zcontours_QL}
\end{figure}

\begin{figure}[h!]
\centering 
    \includegraphics[width=0.4\textwidth]{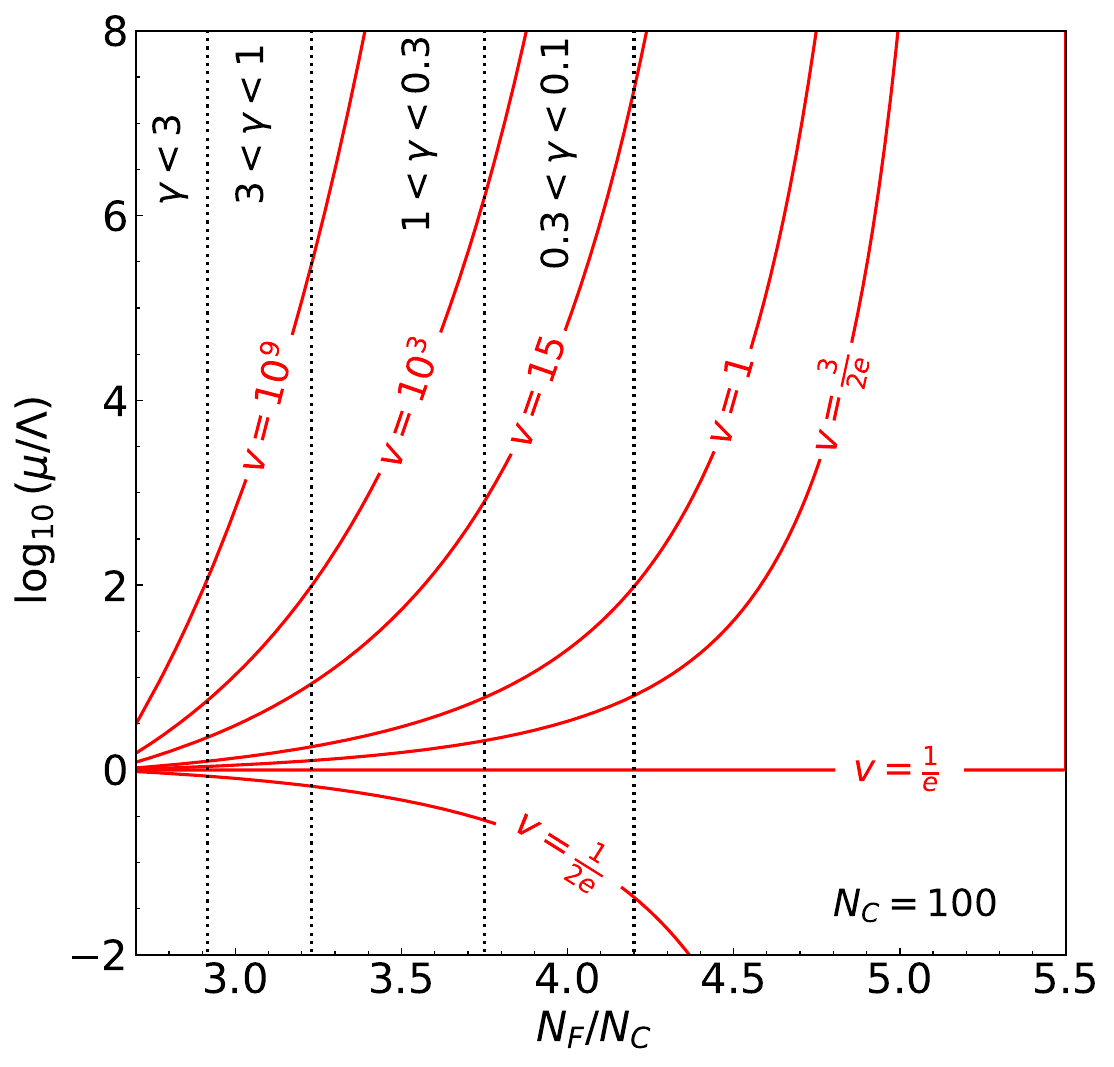}
    \includegraphics[width=0.4\textwidth]{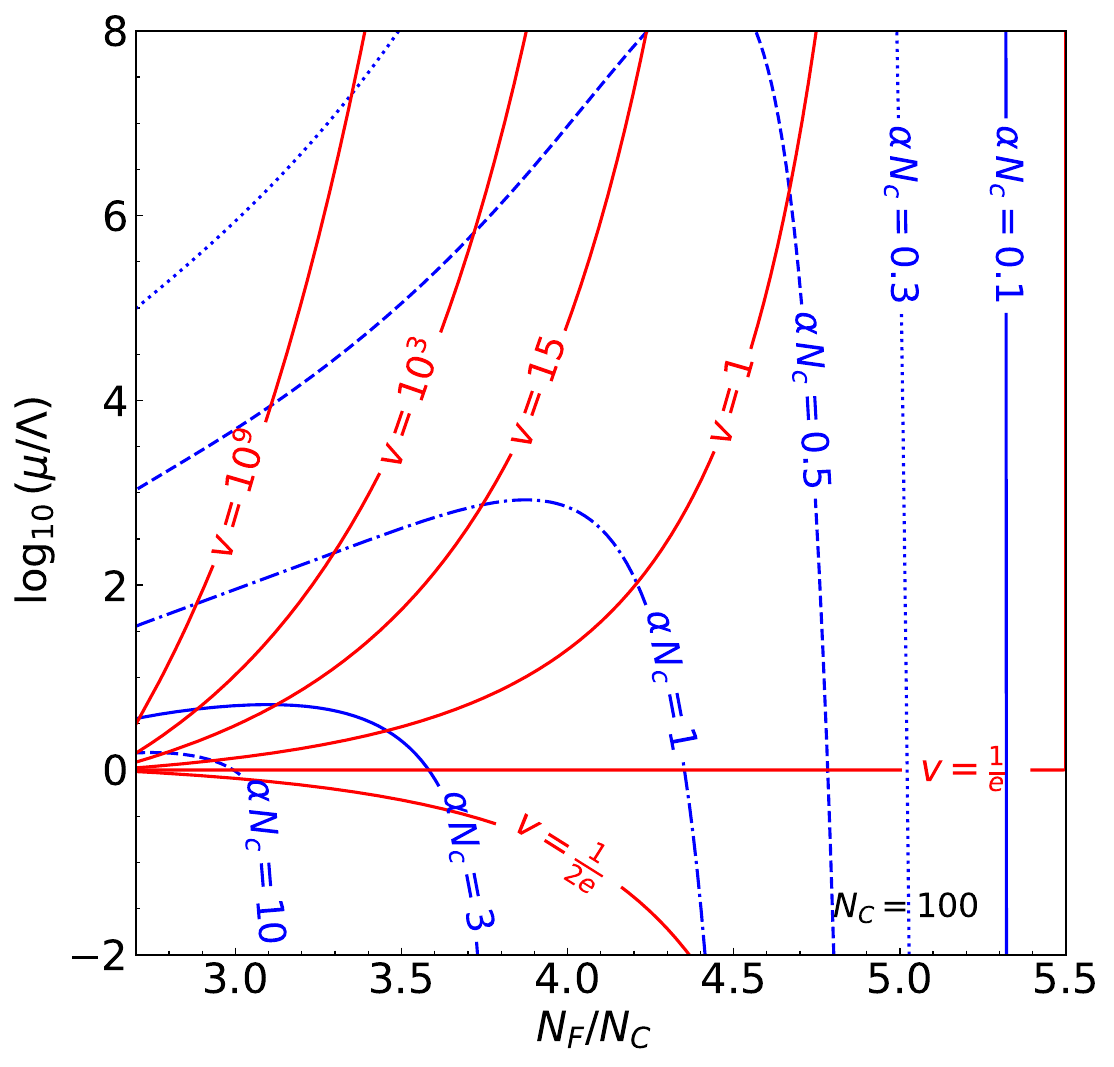}
    \caption{As in fig.~\ref{fig:zcontours_QL}, but for the CW region. Note the definition of $v$ in eq.~\eqref{eq:LamW_argument}.}
    \label{fig:zcontours}
\end{figure}

The nontrivial relations between $v$, $\gamma$, and $\alpha \nc$ as a function of $\fc$ are shown in  fig.~\ref{fig:zcontours_QL} for the QL region and fig.~\ref{fig:zcontours} for the CW region.  It is useful to compare these figures with fig.~\ref{fig:gamma_alphastar}. 

In the QL regime, the relations are rather simple.  The scale $\mu=\Lambda$ corresponds to $v=e$, and as $\ml$ grows exponentially, so does $v$, while $\alpha \nc$ gradually shrinks.  This is familiar from the one-loop behavior of real-world QCD.  The analytically-continued anomalous dimension $\gamma$ is less than $-1$ and varies slowly until $\fc$ is very close to $(\fc)_{CW}$.

In the CW region, the situation is very different. The IRFP occurs at $v\to 0$ with the crossover at $v=1/e$. At the left of the plot, far from the BZ region, we again see exponential growth of $\ml$ is accompanied by exponential growth of $v$, as we did in the QL region.  But the rate of that growth has much stronger $\fc$ dependence than in fig.~\ref{fig:zcontours_QL}, as reflected in the more dramatic change of $\gamma$ with $\fc$.  For $\fc \gtrsim 4$, where $0<\gamma \ll 1$, exponential variation in $v$ is no longer seen, and instead a large range in $\ml$ is compressed to a small range in $v$.  Correspondingly, the coupling barely runs, even in the crossover region. 

The complexity of the CW regime and its difference from the QL regime have an impact on the physics of these models and on the technical question of how to calculate $\alpha(\mu)$.  Recalling that the ratio $\alpha/\alpha_*$ is a function only of $v$, we can see that the domain of $v$ where the Lambert function must be evaluated can vary widely across the $(\fc,\ml)$ plane, especially in the CW case.   Approximations that may work for some portions of this plane will not work in others. The contours of $v$ and $\gamma$ will be useful when we consider possible expansion parameters for various approximations. The contours of $v$ and $\ad \nc$, on the other hand, clarify where perturbation theory is and is not reliable.

%
\subsection{Approximations to the exact solutions}
\label{sec:approximate_solutions}

We now consider various approximations to the exact two-loop coupling that are valid in different ranges of $z$. In particular, we will see how the PDG formula is obtained from the exact expression --- it is not itself an expansion of the Lambert function --- and why it is not useful in most of the CW region.  We will see that though the QL regime can be described using a single expansion of  the Lambert function, this is not possible in the CW regime.  On the other hand, in some parts of the $(\fc,\ml$) plane, certain approximations (not including the PDG formula) may be technically useful for rapid evaluation of the running coupling.

\subsubsection{Ultraviolet expansions}

We first approximate eqn.~\eqref{eq:runnWfn_Lambda}, using  the well-known expansion of the Lambert function \cite{Corless:1996} in polynomials of $\ln \ln v$ divided by powers of $\ln v$, obtaining
\begin{equation}
    \frac{|\ad_*|}{\ad} 
    \approx \mp 1 + \ln v \pm \ln \ln v + \frac{\ln \ln v}{\ln v}
    +{\rm order}\left(\frac{(\ln\ln v)^2,(\ln\ln v)}{(\ln v)^2}\right) \ .
    \label{eq:expandW}
\end{equation}
Here the upper (lower) signs apply for the QL (CW) region.

At large $v$, the leading term in this expression is $\ln v$, and the expansion is in powers of 1/$\ln v$, and so by dropping the extra terms we obtain an approximation valid to third order in the expansion: 
\begin{equation}    
	\frac{|\ad_*|}{\ad}  \approx \mp 1 + \ln v \pm \ln \ln v + \frac{\ln \ln v}{\ln v} \ .
	\label{eq:3OA}
\end{equation}
This formula, which we will refer to as the ``Third-Order Approximation'' (3OA), will be useful below.  Writing it in more familiar terms gives us, 
\begin{equation}
    \frac{1}{\ad} = {\beta_0}\ln\left(\frac{\mu^2}{\Lambda^2}\right) - \frac{1}{\ad_*}\ln\left(\pm\left[1 - {\beta_0\ad_*}\ln\left(\frac{\mu^2}{\Lambda^2}\right)\right]\right)\left(1 + \frac{1}{1 - \beta_0\ad_*\ln\left(\frac{\mu^2}{\Lambda^2}\right)}\right)\ .
    \label{eq:third_order_approximation}
\end{equation}
where the upper (lower) sign is for the QL (CW) region.  This expansion will be valid where $v$ is large; see figs.~\ref{fig:zcontours_QL} and \ref{fig:zcontours}.  Any sign of the IRFP in the CW region, which occurs as $v\to 0$, is now lost; instead this expression diverges at $v=1$, which is at $\mu<\Lambda$ in the QL region but at $\mu=e^{1/2\gamma}\Lambda>\Lambda$ in the CW region. 

To recover the PDG formula, we first take the reciprocal of this expression, expand in $1/(\ln v \mp 1)$, and work only to second-order in the expansion.  This gives us our ``Second-Order Approximation'' (2OA):
\begin{equation}    
	\frac{\ad}{|\ad_*|} =  \frac{1}{\ln v \mp 1}\left(1 \mp \frac{\ln \ln v}{\ln v \mp 1}\right).
	\label{eq:2OA}
\end{equation}

At this point we are still just expanding the Lambert function itself. However, the PDG formula is not merely a function of $v$, and is instead a function of $v$ and $\ml$ obtained as follows.  

We have already assumed $v\gg\ln v\gg 1$ in obtaining eqn.~\eqref{eq:3OA}.  Recalling the sign of $\gamma$, note that
\begin{equation}
	1\ll \ln v = \mp [ \gamma \ln(\mu^2/\Lambda^2) - 1] \approx \mp \gamma \ln(\mu^2/\Lambda^2) = |\gamma| \ln(\mu^2/\Lambda^2)\ . 
	\label{eq:lngammaapprox0}
\end{equation}
Now, if
\begin{equation}
    \Big|\ln(|\gamma|)\Big|\ll\ln \ln (\mu^2/\Lambda^2) \ ,\label{eq:PDGrequirement}
\end{equation}
(that is, if $|\gamma|$ is neither too large nor too small), then we may write  the logarithm of eqn.~\eqref{eq:lngammaapprox0} as
\begin{equation} 
	0 < \ln  \ln v \approx \ln(|\gamma|) +\ln \ln(\mu^2/\Lambda^2) \approx \ln \ln(\mu^2/\Lambda^2) \ .\  \label{eq:lngammaapprox}
\end{equation}
In this case we obtain the PDG formula from eqn.~\eqref{eq:2OA}:
\begin{equation}    
	\frac{\ad}{|\ad_*|} =  \frac{1}{\ln v \mp 1}\left(1 \mp \frac{\ln \ln (\mu^2/\Lambda^2)}{\ln v \mp 1}\right) \  \ . 
	\label{eq:PDGv}
\end{equation}
The upper sign is for the QL region, leading to a divergence in the denominator at $v=e$ ({\rm i.e.} $z=1/e$), where $\mu=\Lambda$.  The lower sign is for the CW case, and gives no such divergence; however, these is still a sub-leading divergence in the numerator at $\mu=\Lambda$. 

In fact, eqn.~\eqref{eq:PDGrequirement} is sufficient but not necessary. The PDG formula also holds if, within the parentheses of eqn.~\eqref{eq:2OA}, the contribution of $\ln |\gamma|/(\ln v\mp 1)$ can be ignored relative to the initial 1.  Thus the condition for the PDG formula to hold (aside from $\ln v\gg 1$) is actually
\begin{equation}
    |(\ln|\gamma|)|\ll {\ln v\mp1} \approx  {\ln v}
    \label{eq:PDGrequirement2}
\end{equation}
or
\begin{equation}
  \Bigg| \frac{(\ln|\gamma|)|}{\gamma}\Bigg|\ll \ln (\mu^2/\Lambda^2) \ .
    \label{eq:PDGrequirement3}
\end{equation}
Assuming $\ln v \gg 1$, eqn.~\eqref{eq:PDGrequirement} implies eqn.~\eqref{eq:PDGrequirement3}, but not the other way around. 

Converting $v$ and $\gamma$ to standard notation, we find the PDG formula takes the same form both in the QL region and the CW region:
\begin{equation}
    \ad(\mu^2)= \frac{1}{\beta_0 \ln(\mu^2/\Lambda^2)}\left(1 - \frac{\beta_1}{\beta_0^2}\frac{\ln[\ln(\mu^2/\Lambda^2)]}{\ln(\mu^2/\Lambda^2)} \right)\ .
    \label{eq:PDG}
\end{equation}
However, we have noted that this formula is only valid when $\ln v\gg 1$ and eqn.~\eqref{eq:PDGrequirement3} holds true, conditions which have completely  different character in the two regions, as is clear from the left panels of figs.~\ref{fig:zcontours_QL} and \ref{fig:zcontours}.  We will explore its range of validity further in a moment.

This process of approximation, moving from the exact two-loop result to the 3OA, the 2OA, and finally the PDG formula, would be expected to be a stepwise progression of decreasing accuracy.  However, some numerical accidents, mainly involving cancellations between higher logarithmic terms, make certain approximations better than they have a right to be.  For SM QCD with $\fc\sim 1$, the PDG formula is just as accurate as the 3OA and more accurate than the 2OA, giving a practical justification for dropping the $\ln|\gamma|$ terms.  It is therefore sufficient for precision QCD applications. 

For theories with other values of $\fc$, this is not always true. Even in the QL region for $\fc > 2$, where $\ln|\gamma|$ is becoming larger, the PDG is less accurate than the 2OA and considerably less than the 3OA.  As for the CW region, not only the PDG but also the 3OA often fail badly, even for $\mu\gg\Lambda$. We will examine these details in a moment, after we discuss two other approximations.

\subsubsection{Infrared and transitional expansions}

In the QL region, we cannot take $\mu<\Lambda$, but in the CW region, the IRFPs are reached as $\mu \to 0$, which is also  $v\to 0$. The small $v$ expansion takes the form
\begin{equation}
    \frac{|\ad_*|}{\ad} = (1 + v - v^2 + \cdots) \ .
    \label{eq:z_equal__expansion}
\end{equation}
Recalling that $v\sim (\mu/\Lambda)^{2\gamma}$, we see that indeed the approach to the fixed point is a power law with exponent $2\gamma$.

\begin{figure}[h!]
\centering 
    \includegraphics[width=0.45\textwidth]{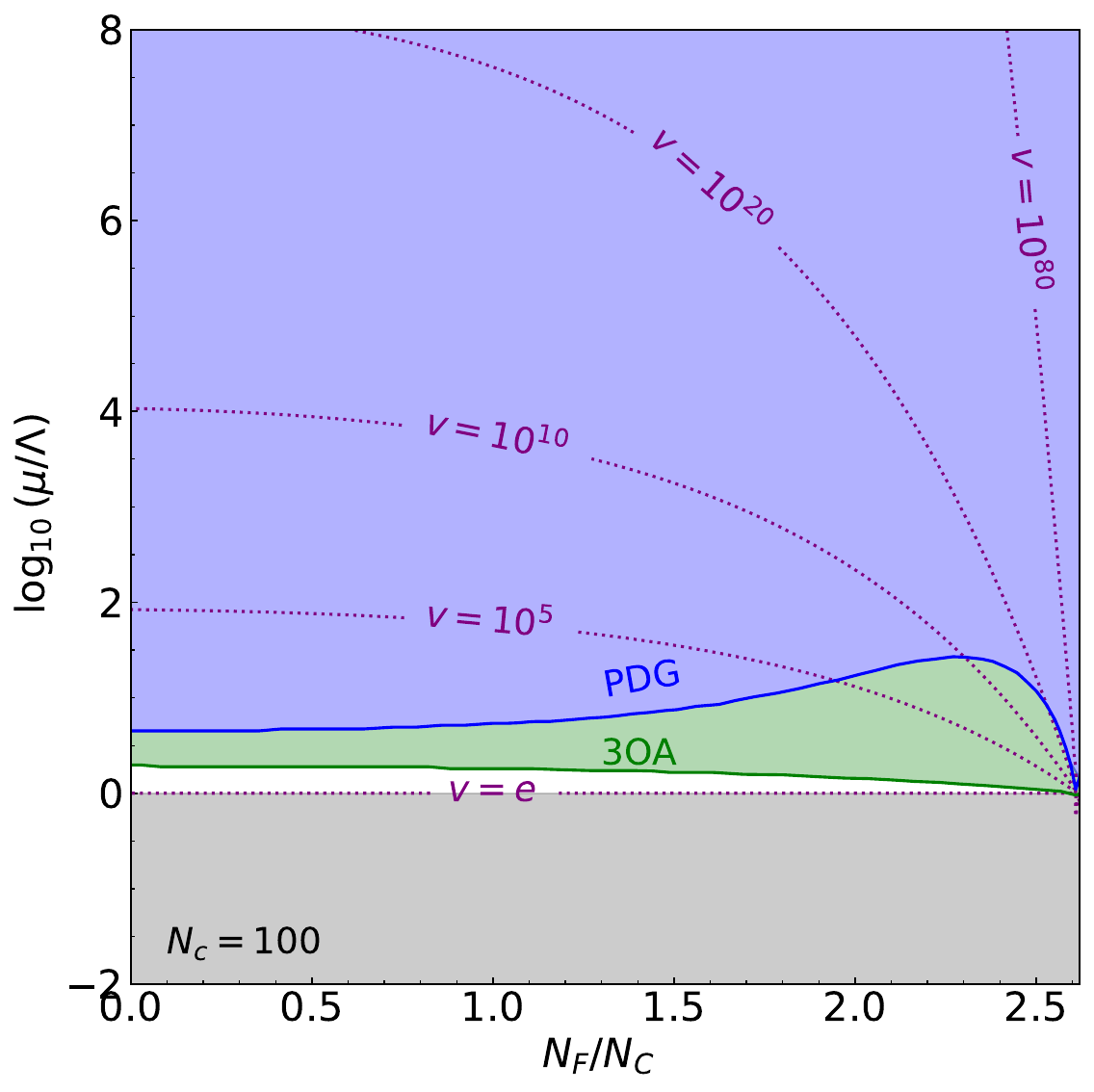}
    \includegraphics[width=0.45\textwidth]{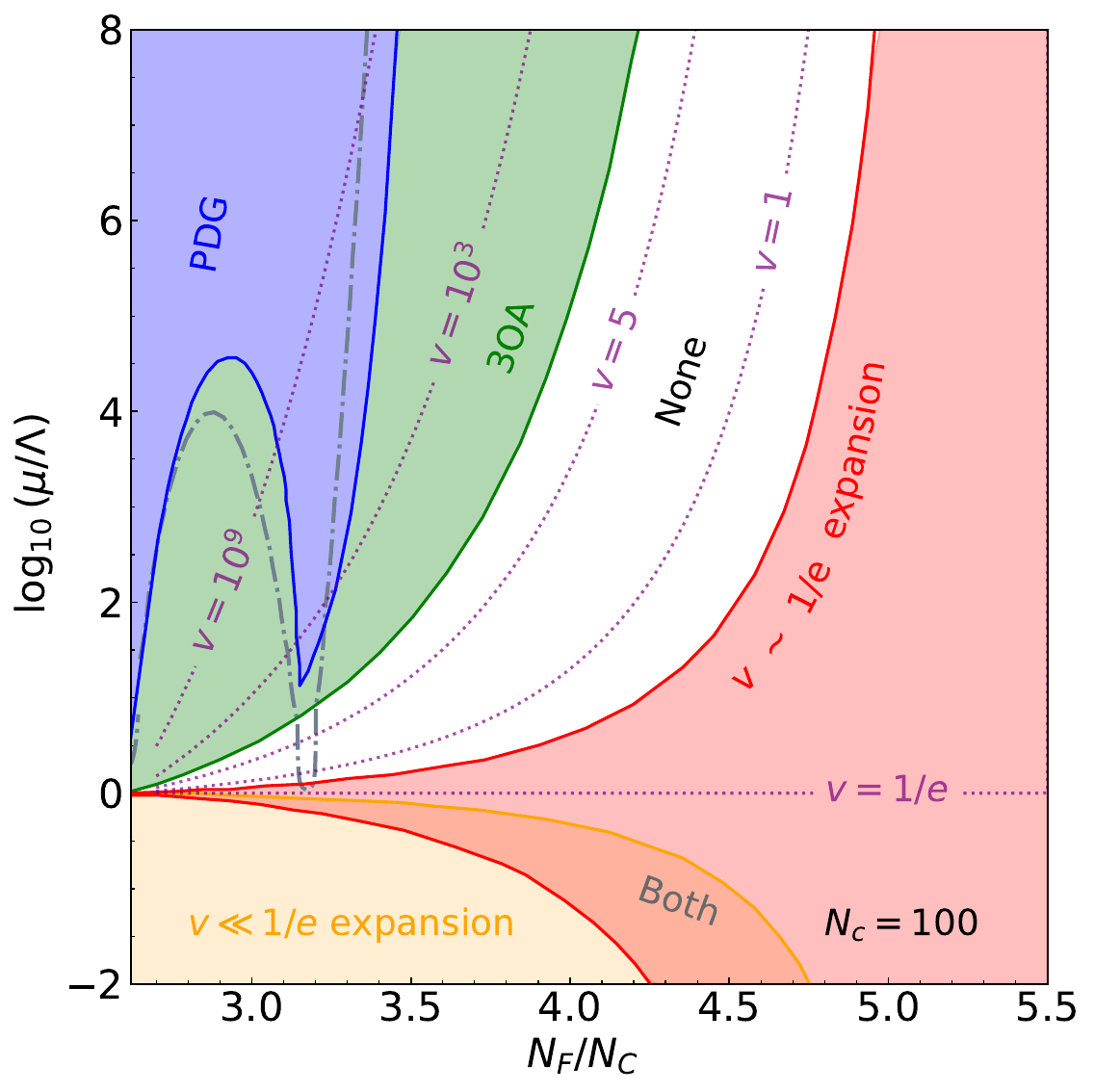} 
    \caption{Validity of approximations to the exact two-loop coupling in the QL region (left panel) and CW region (right panel), as measured by whether they differ from the exact expression by $\geq$ 2\%. In the QL region, the PDG formula only breaks down near $\ml=1$, where the 3OA formula is a minor improvement.  In the CW region, the PDG formula is usually invalid because it requires $\ln \gamma \ll \ln v + 1$ (see eq.~\eqref{eq:PDGrequirement3}; the equation  $\ln \gamma = 0.02( \ln v + 1)$ is shown as a dot-dashed line.) The 3OA applies for $\ln v\gg 1$, but this is not true  in much of the plane. Both small-$v$ and $v \to 1/e$ expansions are valid in the region labeled `Both'. } \label{fig:summary_plot_2_percent}
\end{figure}

The regions of validity of the small $v$ (IR) and large $v$ (UV) expansions of the Lambert function do not overlap.  As a purely technical matter, one might try to combine them with an expansion in the transition region, for $\mu\sim \Lambda$ (and thus $v\sim 1/e$), in hopes that patching these three approximations together might allow one to avoid computing the full Lambert function, thus speeding up parton shower codes. To second order, we have
\begin{equation}
    \frac{|\ad_*|}{\ad} = \left(s_0 +s_1\left(v - 1/e\right) + s_2 \left(v - 1/e\right)^2 + \cdots\right)
    \label{eq:BZPF_expansion}
\end{equation}
where
\begin{eqnarray}
    s_0 &=& 1 + W_0(1/e) \approx 1.28 \ , \ 
 s_1=   \displaystyle\frac{eW_0(1/e)}{1 + W_0(1/e)} \approx 0.592 \nonumber, \\ 
    s_2 &=& -\displaystyle\displaystyle\frac{e^2W_0(1/e)}{2(1 + W_0(1/e))}\left(1 - \displaystyle\frac{1}{\left(1 + W_0(1/e)\right)^2}\right) \approx -0.312. 
\end{eqnarray}
But even this formula's region of validity does not overlap with the 3OA.   This is shown in fig.~\ref{fig:summary_plot_2_percent}, which we now discuss in detail.

%

\subsection{Summary and discussion of approximations}

Fig.~\ref{fig:summary_plot_2_percent} summarizes the validity of these different approximations. The colored regions indicate where each approximation is valid to within 2\% of the exact formula.  The 3OA approximation is valid wherever the PDG formula is valid.  The $v\ll 1/e$ and $v\sim 1/e$ expansions have some overlap, marked ``Both''. 

The two panels are strikingly different.  The ultraviolet 3OA expansion covers almost the entire QL region, with the PDG formula accurate across most of it.  For $\fc\sim 1$, the real-world case, the PDG formula is highly accurate until $\mu< 3 \Lambda_{QCD} \sim 1$ GeV.  But the situation in the CW regime is far less satisfactory.  The 3OA expansion covers only the upper left of the $(\fc,\ml)$ plane, and the PDG formula's region of validity is even smaller.  The infrared $v\ll 1/e$ expansion covers the lower left, capturing the approach to the IRFP.  The transitional $v\sim 1/e$ expansion covers the BZ regime at far right, where the crossover region becomes a very large range of $\ml$ due to the slow running of the coupling.  And yet, even these three approximations do not cover the $v\sim 1$ domain.  (This remains true even if one improves them with higher-order terms in the expansion.)  

The breakdown of the PDG formula within the CW region is easy to understand.  The  unusual shape of the purple region's edge is nearly congruent with the contour
\begin{equation}
	\Bigg|\frac{\ln |\gamma|}{\gamma}\Bigg| = 0.02 \ln(\mu^2/\Lambda^2)\ ,
\end{equation}
which is motivated by eqn.~\eqref{eq:PDGrequirement3} and indicated by a dot-dashed line on the plot. (See also fig.~\ref{fig:zcontours}.)

\begin{figure}[h!]
\centering 
    \includegraphics[width=0.45\textwidth]{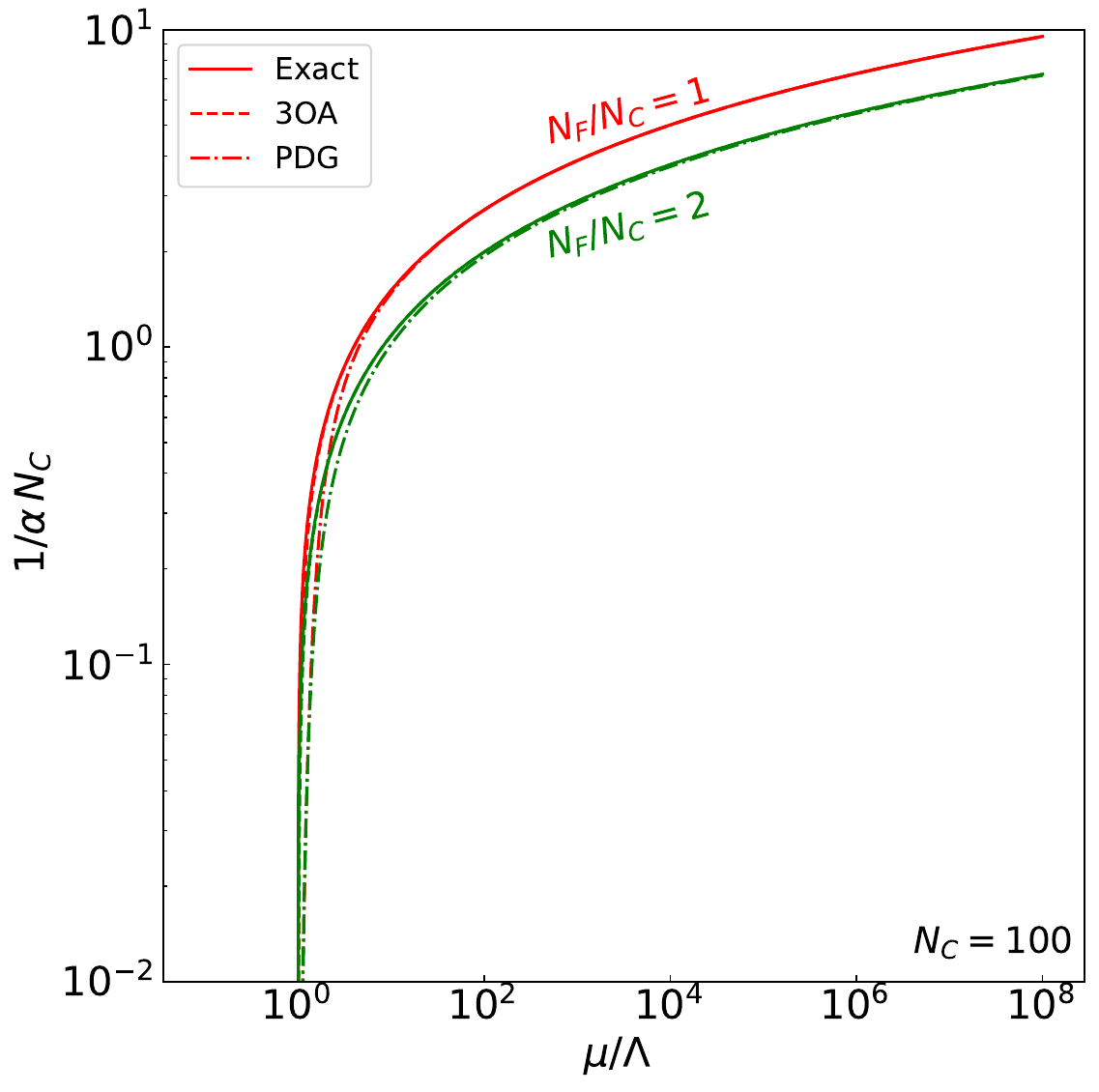}
    \includegraphics[width=0.45\textwidth]{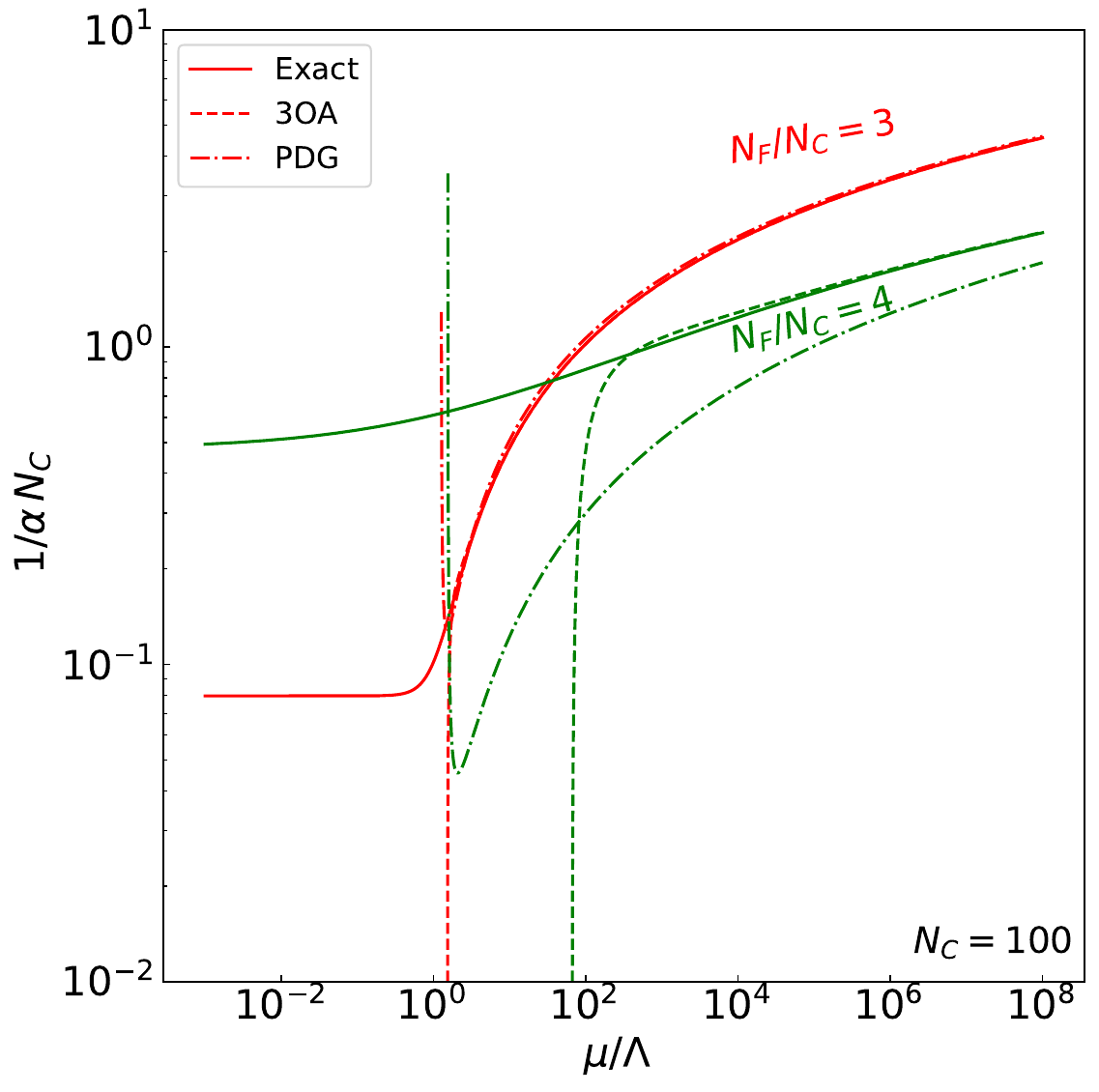} \\
    \caption{Comparison of approximations  for fixed $\fc$ as a function of $\ml$ for the QL (left panel) and CW (right panel) regions. In the CW region for large $\fc$, the PDG formula is especially problematic; not only does it diverge at $\ml=1$, it fails to match the exact solution for large ranges of $\ml>1$. } 
    \label{fig:comparison_ml_space}
\end{figure}

To further clarify the validity of the various UV approximations, we compare them  in fig.~\ref{fig:comparison_ml_space}, showing them as a function of $\ml$ for two different values of $\fc$ in both the QL and CW regions. In the QL region for $\fc \leq 2$, all approximations considered in this work are valid over a large range of $\ml$ and reproduce the exact solution to  good accuracy. In the CW region, the situation is markedly different. For $\fc\sim 3$, the PDG and 3OA solutions  reproduce the exact solution up to $\ml \sim 1$, below which the exact solution develops a fixed point not seen in the UV approximations.  But for $\fc \sim 4$ and larger, the PDG formula, which requires both $\ln v\gg 1$ and  eqn.~\eqref{eq:PDGrequirement3}, fails to reproduce the exact solution even for rather large $\ml$.

\begin{figure}[h!]
\centering 
    \includegraphics[width=0.45\textwidth]{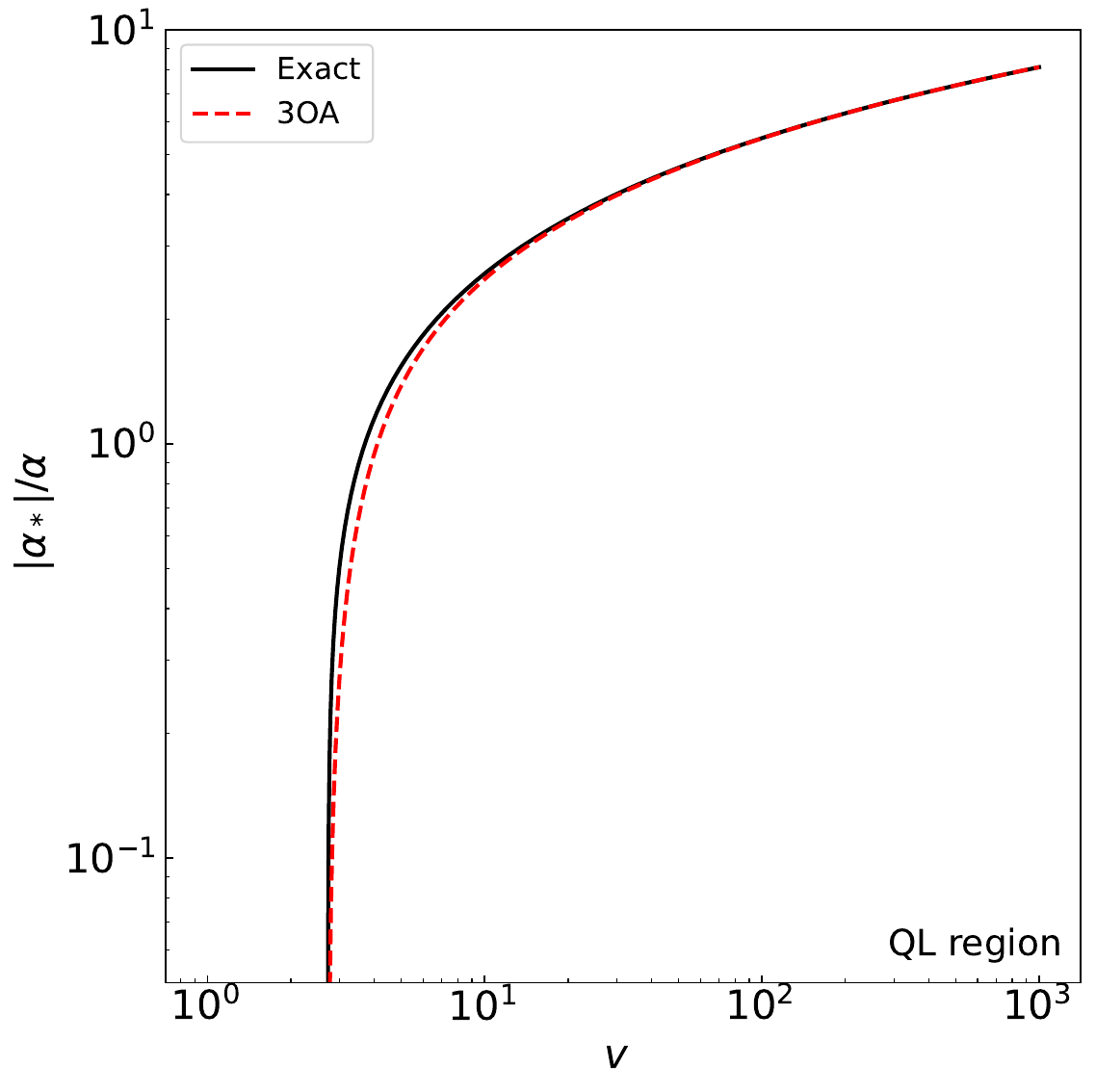}
    \includegraphics[width=0.45\textwidth]{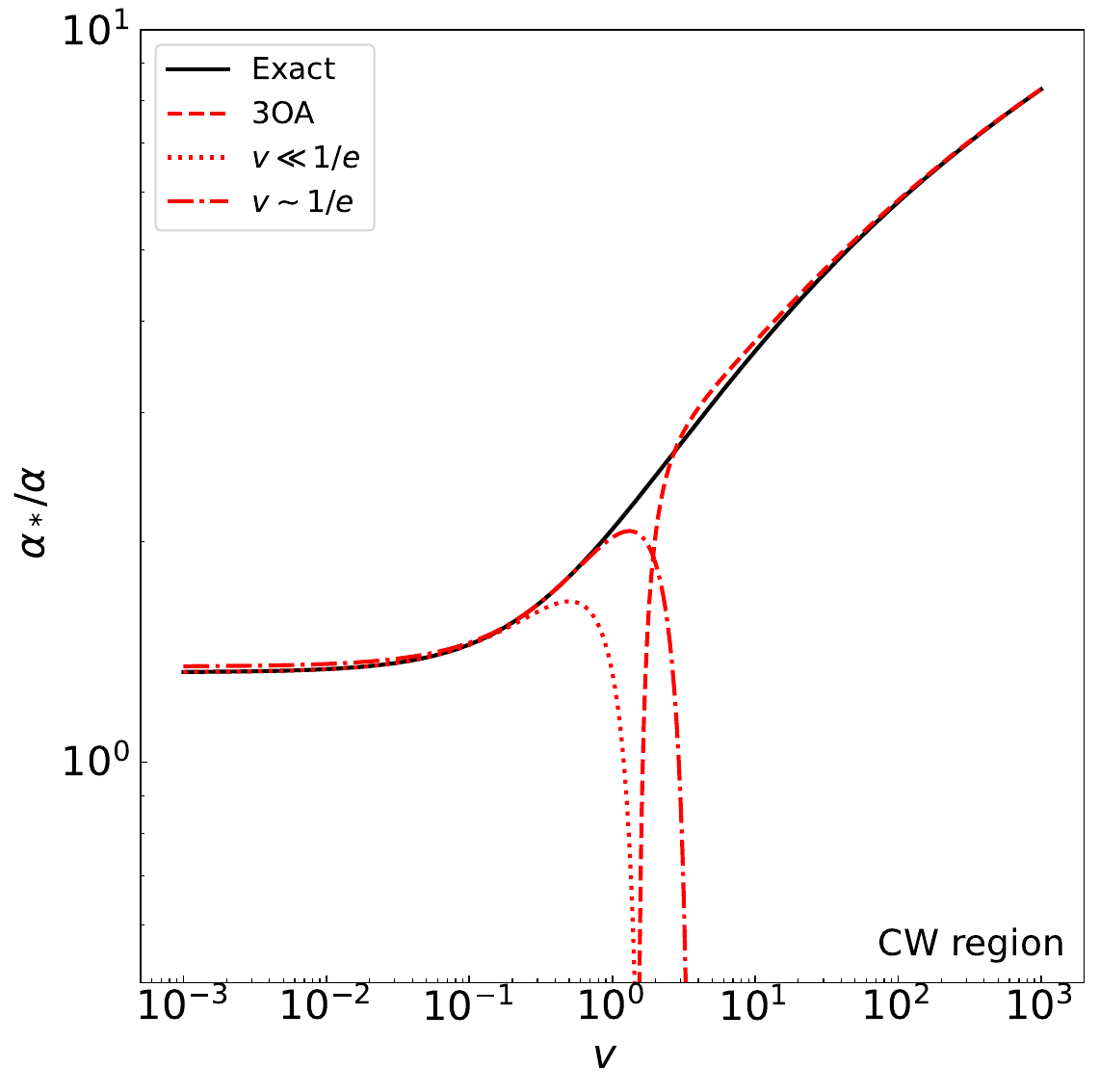} 
    \caption{More detailed comparison of approximate expressions to the exact running coupling, as a function of $v$, for the QL (left panel) and CW (right panel) regimes; at large $\nc$, the curves are independent of $\fc$ within each regime. Note the very different $v$-axis ranges on the two panels. } 
    \label{fig:comparison_z_space}
\end{figure}

Finally, in fig.~\ref{fig:comparison_z_space}, we compare the three approximations given in eqns.~\eqref{eq:3OA}, 
 \eqref{eq:z_equal__expansion} and \eqref{eq:BZPF_expansion} as a function of $v$;
 we exclude the PDG formula, which is not a function of $v$ alone.  The left (right) panel shows the comparison for the QL (CW) regions, independent of $\fc$.  In the QL regime, the 3OA always well-approximates the exact formula except very close to $v=e$ where the coupling diverges. In the CW regime, neither UV nor IR approximations cover the region $v\sim 1$. 

We conclude that for the purposes of HV/DS studies with low-mass dark quarks, the PDG formula of eqn.~\eqref{eq:PDG} may and should be dropped; anywhere it holds it can be replaced by the 3OA formula of eqn.~\eqref{eq:third_order_approximation}, which is no more difficult to compute.  (Note that mass thresholds from heavier dark quarks do complicate this statement, but we see no obstruction to using the full Lambert function across such thresholds.) By improving the 3OA with higher-order terms, one can obtain even more accurate approximations to the Lambert function in the UV, almost completely covering the QL region and further extending its validity in the CW region. The infrared and transitional expansions for the CW can similarly be expanded to higher orders if needed, extending their range.  A gap near $v\sim 1$ will still remain, but it is not technically difficult to close it because the Lambert function is relatively gentle there.  A lookup table could be used across the gap, or one could use a combination of additional high-order expansions, such as those suggested in~\cite{Veberic:2010ay,Veberic:2012ax}; even simple Taylor series around $v=1$ and $v=3$ seem sufficient.  We leave the optimization of this choice to future work, but see no practical obstacle.

%

\section{The Sudakov veto algorithm at two-loops}
\label{sec:two_loop_sudakov}

The parton shower encoded in most current event generators is based on leading order parton splitting functions. (For discussion of higher-order parton showers, see, e.g., \cite{Dasgupta:2020fwr,vanBeekveld:2023ivn, panscalesweb,Loschner:2021keu,Platzer:2022jny}). This shower is usually combined with the one-loop running coupling, but most generators also give the user the option to use an approximation to the two-loop running coupling.  

In PYTHIA the two-loop coupling is approximated by the PDG formula given in eqn.~\eqref{eq:PDG}.  As we have seen in section~\ref{sec:approximate_solutions}, this does not allow full exploration of the CW regime, where the exact two-loop running coupling should be used instead to allow exploration of the full showering phase space.

One might question whether the combination of the exact two-loop running coupling with a parton shower based on leading-order splitting functions is a consistent approximation, and in what settings.  We leave this question for future consideration, but note three facts.   First, in the short term, phenomenological studies of and searches for confining hidden valleys/dark sectors are subject to substantial experimental uncertainties from complex backgrounds and theoretical uncertainties that can arise from hadronization effects. A parton shower need not have high precision to be useful, as long as it is not too inaccurate. Second, PYTHIA's shower is reasonably well justified both for a constant coupling and for a one-loop coupling.  The two-loop running coupling in the CW regime lies between these two cases, running slower than the latter in the UV and approaching the former in the IR.  Third, as a practical matter, the combination of the two-loop coupling with the existing PYTHIA parton shower is the quickest path to initial studies of dark showering in the CW regime.  While improvements over this approach will be welcome, the immediate need is to make such studies possible in the first place.

For PYTHIA or other similar generators to produce a complete parton shower in the CW, necessary modifications involve not only the coupling but also the Sudakov factor  $\Delta_a(Q_1^2,Q_2^2)$, which enters in the time-like evolution central in modeling final state radiation (FSR).  This factor represents the probability that, once an emission of a parton has occurred at a scale $Q_1$, no emission takes place between $Q_1$ and a lower scale $Q_2$.  As we will see in a moment, PYTHIA's current strategy for computing the Sudakov factor for a two-loop coupling, which works well in the QL region, does not work in the CW regime.  We will show how this obstacle can easily be evaded.\footnote{The following is specifically applicable to the Hidden Valley module of PYTHIA 8, but can be straightforwardly generalized for other event generators. For instance, for QCD simulations, Herwig uses the ExSample library~\cite{Platzer:2011dr}  for sampling the Sudakov factor. While this method can handle the two-loop coupling in the CW regime, our results below are nevertheless applicable there as well.} 

Consider a parton of type $a$ at a given stage in the parton showering process. By evaluating the Sudakov factor, we can consider how this parton evolves from its emitted scale $Q_1^2$  to the lower scale $Q_2^2$ where it undergoes branching. At this scale, we can sample a specific branching process $a\to bc$ and select the energy fractions $\xi, 1-\xi$ of the daughter partons $b,c$. We repeat this procedure for partons $b$ and $c$ and so forth until the infrared cutoff on the shower is reached.  

The Sudakov factor for the parton $a$ not to branch between scales $Q_1$ and $Q_2$ is given by 
\begin{eqnarray}
    \Delta_{a}\left(Q_2^2,Q_1^2\right) = \exp\left(-\int_{Q^2_2}^{Q_1^2} \frac{{dQ^\prime}^2}{{Q^\prime}^2}\frac{\ad({Q^\prime}^2)}{2\pi} \int_{\xi_{min}({Q^\prime}^2)}^{\xi_{max}({Q^\prime}^2)} \sum_{b,c}P_{a\to bc}(\xi^\prime) d\xi^\prime \right) \ ,
    \label{eq:sudakov_general}
\end{eqnarray}
where $P_{a\rightarrow bc}$ are the standard Altarelli-Parisi (AP) splitting functions.
All possible branchings of the initial parton, $a\to bc$, are summed over. The variable $\xi'$ is the fraction of energy given to parton \textit{b} (with $1-\xi'$ given to parton \textit{c}); the boundaries of integration $\xi_{min}({Q^\prime}^2), \xi_{max}({Q^\prime}^2)$ are determined by the kinematics of the branching~\cite{Sudakov:1954sw}.   

In a Monte Carlo event generator, every step of parton splitting requires the generation of a new set of $\left[Q^2,\xi\right]$ according to the probability given in eq.~\eqref{eq:sudakov_general}. This task is usually achieved by means of the veto algorithm~\cite{Sjostrand:1987su,Ellis:1996mzs,Sjostrand:2006za,Buckley:2011ms}, see also~\cite{Lonnblad:2012hz,Platzer:2011dr,Platzer:2011dq}. In what follows, we focus our attention only on the selection of $Q^2_2$, as the strategy for selecting the next value of $\xi$  requires no changes. 

In the veto algorithm, the Sudakov integrand is overestimated and simplified by replacing the splitting functions $P(\xi)$ with overestimates $\tilde P(\xi)$ and by expanding the integration region, where the boundaries $\tilde\xi_{min}$ and $\tilde\xi_{max}$ of the $\xi$ integral are chosen to be independent of $Q'$ and $Q_2$ (though not necessarily of $Q_1,$ the initial $Q^2$ of emission.) 
Defining
\begin{equation}
	\epa(\tilde\xi_{min},\tilde\xi_{max})\ = \sum_{b,c} \int_{\tilde{\xi}_{min}}^{\tilde{\xi}_{max}} \tilde{P}_{a\to bc}(\xi^\prime) d\xi^\prime
\end{equation}
and
\begin{equation}
	\delaepa(Q_1^2,Q_2^2)\equiv 
	\exp\left(-\int_{Q^2_2}^{Q_1^2}\frac{\ad(Q'^2)}{Q'^2} dQ'^2 \right) = \exp\left(-\int_{\ad(Q_2^2)}^{\ad(Q_1^2)}\frac{\ad^\prime}{\beta(\ad^\prime)} d\ad^\prime\right) \ ,
	\label{eq:kappadefn}
\end{equation}
we may write the Sudakov factor with the overestimated integrand as $\dela$, with 
\begin{eqnarray}
    \dela\left(Q_2^2,Q_1^2\right) &=& \delaepa(Q_1^2,Q_2^2)^{\epa(\tilde\xi_{min},\tilde\xi_{max})/2\pi}.
    \label{eq:sudakov_kappa_delaepa}
\end{eqnarray}
Recall that $\tilde\xi_{min},\ \tilde\xi_{max}$  are independent of $Q_2$ by construction. 

To obtain a value of $Q_2$ for the next branching, given the initial scale $Q_1^2$, one takes a random number $R$ between 0 and 1 and solves $\dela = R$ for $Q_2$ as a function of $Q_1$. The solution for constant coupling or one-loop running coupling is well-known \cite{Webber:1983if}: 
\begin{equation}
    Q_2^2 = \displaystyle
    \Bigg\{
    \begin{matrix}
    Q_1^2 \times \delaepa^{1/\ad}  \ \ \ \  &({\rm constant \ \ad}) \cr \Lambda^2 \times \left[\frac{Q_1^2}{ \Lambda^2}\right]^{\delaepa^{\beta_0}} \ \ \ \ &({\rm one-loop \ \rm\ad})
    \end{matrix} \ . 
    \label{eq:sudakov_const_a_oneloop}
\end{equation}
where in this equation $\delaepa$ is to be understood not as defined in eqn.~\eqref{eq:kappadefn} but as a function of the random number $R$, namely 
\begin{equation}
    \delaepa = R^{2\pi/\epa} \ ,
    \label{eq:kappaR}
\end{equation}
which still depends on $\tilde\xi_{min},\ \tilde\xi_{max}$ but not on $Q_2$.  A separate $Q_2$-independent procedure that addresses the overestimate of the $d\xi'$ integrand then allows selection of $\xi$  and corrects for the overestimate in $\epa$.

However, when the coupling is taken to run at two-loop order, then PYTHIA evaluates $\delaepa$  using a second veto algorithm.  This is done by using the one-loop coupling as an overestimate of the two-loop coupling within the $d\alpha'$ integral.  Such a strategy works well in the QL, where $\beta_1>0$ and therefore $\alpha_{{\rm 1-loop}}(\ml) > \alpha_{{\rm 2-loop}}(\ml)$ for all $\mu>\Lambda$.  But in the CW regime, $\beta_1$ has the opposite sign.  Worse, $\ad_{{\rm 2-loop}}(\ml)$ must often be evaluated for $\mu<\Lambda$, but $\ad_{{\rm 1-loop}}(\ml)$ is not even defined there. 

Fortunately, the same approach used in eqn.~\eqref{eq:sudakov_const_a_oneloop} can be used here. The integral over the exact two-loop running coupling can be performed, and just as with a one-loop or constant coupling, the equation $\Delta_a=R$ can then be solved for $Q_2$,  eliminating the need for a veto algorithm involving $\ad$. 
Specifically, as we demonstrate in  appendix \ref{app:sudakov}, 
\begin{equation}
Q_2^2 = \Lambda^2 \left(\frac{Q_1^2} {\Lambda^2}\right)^{\delaepa^{\beta_0}}
    \left(\delaepa^{\beta_0}\left[\mp e W_{n}(\mp z_1)\right]^{1-\delaepa^{\beta_0}}\right)^{1/\gamma} \ \ ({\rm two-loop}\ \alpha) \ ,
    \label{eq:sudakov_two_loop_mb}
\end{equation}
where one takes the upper (lower) sign and $n=-1$ ($0$) for the QL (CW) region, $z_1$ is the variable $z$ from equation \eqref{eq:argumentQL} defined at $\mu^2=Q_1^2$, $\gamma$ is the two-loop critical exponent defined in equation \eqref{eq:gamma_twoloop}, and $\delaepa$ is to be understood as in eqn.~\eqref{eq:kappaR}.
Note this expression implicitly depends on $\fc$ through $\beta_0$ and $\gamma$.

With this closed-form expression for $Q_2^2$, applicable in both the QL and CW regimes, the veto algorithm for $\ad$ and its associated problems are sidestepped. Meanwhile the veto algorithm associated with overestimating the $\xi$ integral remains unchanged.

\begin{figure}[h!]
\centering 
    \includegraphics[width=0.45\textwidth]{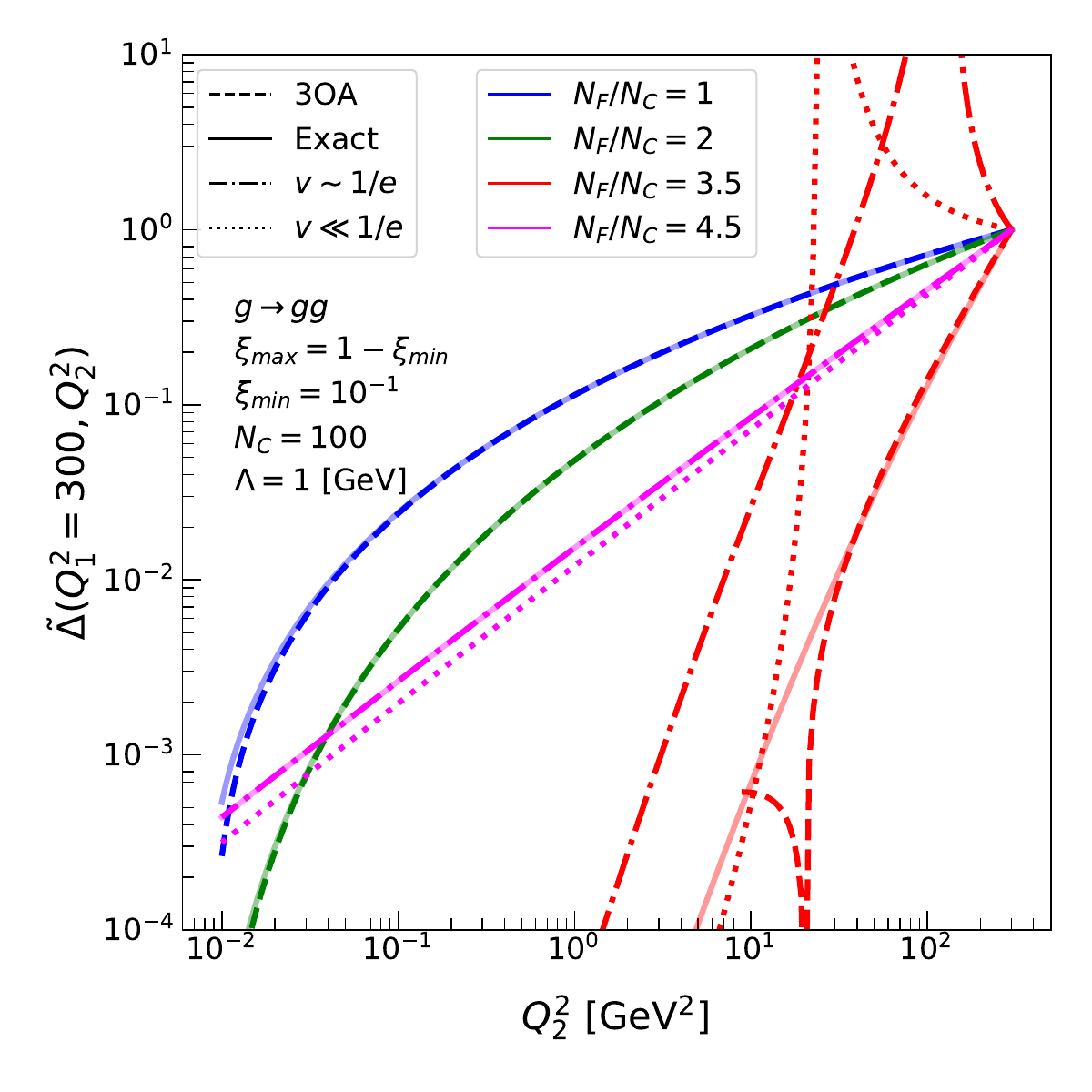}
    \includegraphics[width=0.45\textwidth]{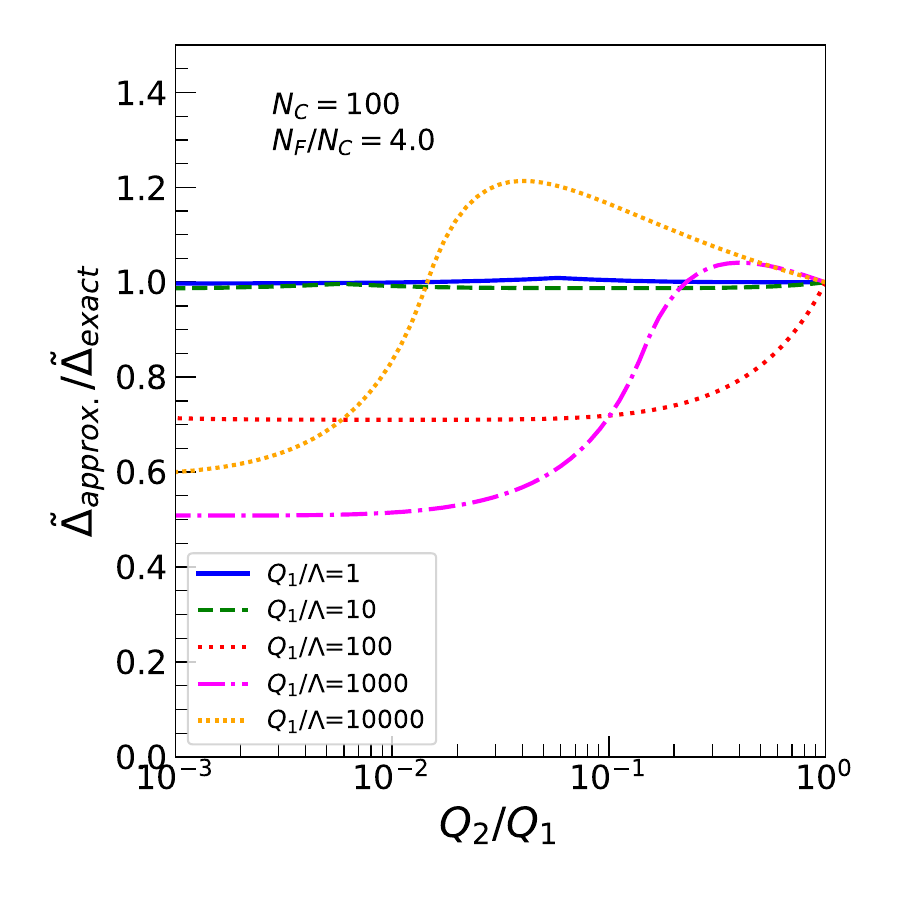} 
    \caption{Estimated Sudakov factor $\tilde\Delta_g$ as given in expression~\eqref{eq:sudakov_kappa_delaepa}, for the $g\to gg$ splitting function only. For various $\fc$ we compare (left panel) the values of $\tilde\Delta$ computed using the exact two-loop $\alpha(\mu    
    )$ to those computed using individual approximations presented in Sec.~\ref{sec:explicit_solutions}. We also show (right panel) the ratio of $\tilde\Delta_g$ computed using an approximate $\alpha(\mu)$, where at each $\mu/\Lambda$ the best available approximation is used, to that computed using the exact two-loop $\alpha(\mu)$.} 
    \label{fig:sudakov}
\end{figure}

As a final illustration of the effect of using approximations, we show in Figure~\ref{fig:sudakov} how they can impact the estimated Sudakov factor. We explicitly compute eqn.~\eqref{eq:sudakov_kappa_delaepa}, where we account only for the $g\to gg$ splitting function with $\xi_{min} = 0.1$. In the left panel, fixing the value of $Q_1^2 = 300$ GeV$^2$ and varying $Q_2^2$, we show the Sudakov factors for several values of $\fc$, comparing the calculations using the exact two-loop coupling to those using the approximate forms discussed in Sec.~\ref{sec:explicit_solutions}.  
For the QL region with $Q_i$ well above $\Lambda$, the 3OA approximation to the exact two-loop $\alpha_s(\mu)$ reproduces $\tilde \Delta_a(Q_2^2,Q_1^2)$  at the percent level. But this is not so in the CW regime, where the 3OA approximation can fare poorly even for $Q_1,Q_2 \gg \Lambda$.  At this value of $Q_1$, the small$-v$ approximations can work for $\fc\gtrsim 4.5$, but not for  $\fc\sim 3.5.$ 


Of course, we need not limit ourselves, when computing the integral in~\eqref{eq:kappadefn}, to just a single approximation to the coupling.
In the right panel, we plot the ratio of approximate to exact Sudakov factors for $\fc=4$, where to calculate the former we choose the best available approximation for the coupling constant $\alpha(\mu)$ at that value of $\mu/\Lambda$.  Specifically, we choose whichever of the 3OA, $v\sim 1/e$ or $v\ll 1/e$ approximations works best for that value of $v$; see Figure~\ref{fig:comparison_z_space} (right).  Even so, the right panel of Figure~\ref{fig:sudakov} shows that for some values of $Q_1,Q_2$ we may still find deviations in $\tilde\Delta_a$ as large as 50\%.  Noting the location of the unshaded region in Figure~\ref{fig:summary_plot_2_percent} (right), we see that the largest deviations occur when $Q_1$ lies in or above that region, while $Q_2$ lies in or below it.

\normalsize
\section{Outlook and discussion}
\label{sec:conclusion}

Field theories with infrared fixed points, such as those similar to QCD but with higher $\fc$,  are conceptually interesting in and of themselves.  But the possibility that HV/DS models of this type might exist, and yet might have escaped detection at the LHC, mandates that we learn, more practically, how to simulate them.

Specifically, the properties of dark jets in HV/DS models in the conformal window can only be understood, even qualitatively, with simulations that can capture the running of the coupling beyond one-loop, including the crossover into the approach to the IRFP.  At present, however, existing generators approximate the two-loop coupling in a way that is insufficient for this purpose.  Not only is the IRFP invisible to such approximations, the ones currently used are not always  accurate in the UV either, as we have seen in fig.~\ref{fig:summary_plot_2_percent}.  The use of exact two-loop RGE solutions, going beyond the well-known approximate PDG formula, is a necessary step.

As we have argued here, this is most directly rectified by replacing the approximate two-loop running coupling by its exact form, eqn.~\eqref{eq:runnWfn_Lambda}, which involves the Lambert function.  While this change is straightforward conceptually, it is not entirely trivial technically, as computation speed must be maintained.   We have addressed the practical challenges of this computation by examining various approximation schemes for this function, showing that simple ultraviolet and infrared expansions are not sufficient, but that a larger set of expansions, possibly combined with a look-up table in small regimes, should be enough.  A new approach to computing the Sudakov factor in the parton shower is also needed; we propose a modification of the standard veto algorithm in eqn.~\eqref{eq:sudakov_two_loop_mb} and the discussion preceding it, which again requires computation of the Lambert function. 

Our analysis is restricted to two loops, and one must wonder what aspects of it might properly represent the actual physics of a real HV/DS model.
The mere use of the full two-loop running coupling  is far from sufficient for accurate results, and a complete NLL parton shower in the CW regime will not soon be available.
In addition, there are higher-order and non-perturbative effects that will be important anywhere outside the BZ region.  As  one moves to lower $\fc$, the fixed point coupling $\ad_*$ and the anomalous dimension $\gamma$ grow, so that two-loop approximations are no longer accurate for $\mu\lesssim\Lambda$.  

However, it is far from clear that going to higher and finite order would add much accuracy or precision.  As noted already in section~\ref{subsec:model_classification}, two-loop approximations to the coupling are already enough to capture the key qualitative features of sectors in the conformal window, namely a crossover from weak-coupling logarithmic running to approximate fixed-point behavior.  There are no known qualitative features that appear at higher orders.  Furthermore, the quantitative benefits of higher orders are limited.  In the BZ regime, higher orders are unneeded, while conversely the loop expansion will be poor once $\fc\sim  4.5$ or below (see fig.~\ref{fig:zcontours}). This leaves a relatively narrow zone in which three-loop corrections could improve the precision of two-loop approximations. On top of this, higher-order corrections to the coupling exhibit strong scheme-dependence, which can only be mitigated by a full and consistent higher-order parton shower that lies far out of reach.

Instead, it may be more important in the near- and medium-term to obtain scheme-independent, fully non-perturbative information from lattice gauge theory.  Even imprecise estimates of how $\gamma$ depends on $\fc$, and the true value of $(\fc)_{CW}$, may prove more valuable for collider searches and their interpretation than quantitative but scheme-dependent information from higher loop corrections to $\alpha(\mu)$.

These challenges notwithstanding, our work represents a first step in the direction of simulating theories in the CW regime. We have seen that the evaluation of the full running coupling at two loops, and a corresponding approach to the evaluation of the Sudakov factor, are prerequisites. In a forthcoming paper, we will discuss the implementation of these methods and will illustrate the associated phenomenology, sketching the effect of our framework and discussing underlying collider signatures. 

\section*{Acknowledgements}
SK and JL are supported by the FWF research group funding FG1 and FWF project number P 36947-N. We thank Reinhard Alkofer, Matteo Cacciari, Einan Gardi, Jack Holguin, Daniel Litim, Axel Maas, Stephen Mrenna, Simon Pl\"atzer, Gavin Salam, Torbj\"orn Sj\"ostrand,  Gregory Soyez, Jesse Thaler and Fabian Zierler
for valuable discussions. SK is grateful to Mainz Institute for Theoretical Physics (MITP) of the Cluster of Excellence PRISMA$^+$ project (Project ID 390831469), for its hospitality and its partial support during the completion of this work. MJS is grateful to Harvard University for its hospitality.
\appendix 

\section{{Overestimation of the Sudakov factor in event generators}}
\label{app:sudakov}
Recalling from section~\ref{sec:two_loop_sudakov},  
the leading-order Sudakov factor $\Delta_a$ for a parton $a$ is
\begin{eqnarray}
    \Delta_{a}\left(Q_2^2,Q_1^2\right) = \exp\left(-\int_{Q^2_2}^{Q^2_1} \frac{{dQ^\prime}^2}{{Q^\prime}^2}\frac{\ad({Q^\prime}^2)}{2\pi} \int_{\xi_{min}({Q^\prime}^2)}^{\xi_{max}({Q^\prime}^2)} \sum_{b,c}P_{a\to bc}(\xi^\prime) d\xi^\prime \right) \ ,
    \label{eq:appendix_sudakov_general}
\end{eqnarray}
where the $P_{a\rightarrow bc}$ are the Altarelli-Parisi (AP) splitting functions.  Overestimating the integrand of $\Delta_a$ allows us to write it in a modified form $\dela$ in terms of overestimated splitting functions $\tilde P$, along with an overestimated integration region whose boundaries  $\xi_{min},\xi_{max}$ are independent of $Q'$ and $Q_2$.  The modified Sudakov factor then can be written
\begin{eqnarray}
    \dela\left(Q_2^2,Q_1^2\right) &=& \delaepa(Q_1^2,Q_2^2)^{\epa(\tilde\xi_{min},\tilde\xi_{max})/2\pi}
    \label{eq:sudakov_generalx}
\end{eqnarray}
where 
\begin{equation}
	\epa(\tilde\xi_{min},\tilde\xi_{max})\ = \sum_{b,c} \int_{\tilde{\xi}_{min}}^{\tilde{\xi}_{max}} \tilde{P}_{a\to bc}(\xi^\prime) d\xi^\prime
\end{equation}
and
\begin{equation}
\delaepa(Q_1^2,Q_2^2)\equiv  \exp\left(-\int_{Q_2^2}^{Q_1^2}\frac{\ad(Q'^2)}{Q'^2} dQ'^2 \right) = \exp\left(-\int_{\ad(Q_2^2)}^{\ad(Q_1^2)}\frac{\ad^\prime}{\beta(\ad^\prime)} d\ad^\prime\right)\ . 
\end{equation}
When we substitute the two-loop $\ad$, either form of the integral can be computed, giving
\begin{equation}
	\delaepa=  \exp\left[\frac{1}{\beta_0}\ln \left(\displaystyle\frac{1 - \displaystyle\frac{\ad_*}{\ad(Q^2_2)}}{1 - \displaystyle\frac{\ad_*}{\ad(Q^2_1)}}\right)\right] = \left[\frac{W_{n}(\mpx z_2)}{W_{n}(\mpx z_1)}\right]^{1/\beta_0}
\end{equation}
where $\mpx=n=-1$ for the QL region and $\mpx=+1$, $n=0$ for the CW region. We have used the definitions of $\ad$ in \eqref{eq:runnWfn_Lambda}, of $z$ in \eqref{eq:argumentCW} and of $v$ in \eqref{eq:LamW_argument}, and the $z_i$ are defined as $z$ at $\mu^2=Q_i^2$.
To solve this for $Q_2^2$, we rewrite this as 
\begin{equation}
	W_{n}(\mpx v_2^{\mpx }) = {\delaepa^{\beta_0}}W_{n}\left(\mpx v_1^{\mpx }\right)\ .
	\label{eq:WW}
\end{equation}
Then we use $x=W(x)\exp[W(x)]$ and eqn.~\eqref{eq:WW} repeatedly, giving
\begin{eqnarray}
	\sigma v_2 &=& W_n(\sigma v_2^\sigma)\exp\left[W_n(\sigma v_2^\sigma)\right] = \delaepa^{\beta_0} (\sigma v_1^\sigma) 
	\exp\left[W_n(\sigma v_2^\sigma)-W_n(\sigma v_1^\sigma)\right]\\
	&=& \delaepa^{\beta_0} (\sigma v_1^\sigma) 
	\exp\left[(\delaepa^{\beta_0}-1) W_n(\sigma v_1^\sigma)\right]
	= \delaepa^{\beta_0} (\sigma v_1^\sigma) 
	\exp\left[\frac{W_n(\sigma v_1^\sigma)}{\sigma v_1^\sigma}\right]^{1-\delaepa^\beta_0}.
\end{eqnarray}
From here and the definition of $v$, we immediately obtain our result of eqn.~\eqref{eq:sudakov_two_loop_mb}:
\begin{equation}
    Q_2^2 = \Lambda^2 \left(\frac{Q_1^2} {\Lambda^2}\right)^{\delaepa^{\beta_0}}
    \left[\delaepa^{\beta_0}\left(\mpx e W_{n}(\mpx v_1^{\mpx })\right)^{1-\delaepa^{\beta_0}}\right]^{1/\gamma}.
    \label{eq:sudakov_two_loop}
\end{equation}
This gives us a closed-form expression that allows us to select the next $Q^2$ for branching without using a veto algorithm for $\ad$. As a check on this formula, we consider two interesting limits.  First, in the CW region, we may take $\ad$ to its fixed point $\ad_*$, which occurs when $\mu/\Lambda \to 0$. In case the argument of the Lambert $W$ function is small and can be approximated within the CW region as $W_0(v)\approx v$. Then \eqref{eq:sudakov_two_loop} can be written as
\begin{equation}
    Q_2^2  = Q_1^2 \times \delaepa^{1/\ad_*}
\end{equation}
which is nothing but the scale relation for $\dela$ with constant $\ad=\ad_*$ as discussed in eqn.~\eqref{eq:sudakov_const_a_oneloop}. Thus our two-loop relation reduces to the expected constant $\ad$ relation in the constant coupling limit. 

Another useful limit involves $\beta_1\to0$, which occurs at $\fc\to\left(\fc\right)_{CW}$ and $\gamma\to\infty$. In this limit the two-loop and one-loop running couplings are the same, and we expect to find the one-loop relation eqn.~\eqref{eq:sudakov_const_a_oneloop}. For convenience, we take the limit from the CW side, using the 0$^{\rm{th}}$ branch of the Lambert $W$ function, which for large $\gamma$ is given by $W_0(v)\approx \ln(v) = \gamma\ln(Q^2/\Lambda^2) - 1$. Using the fact that $(a \gamma + b)^{c/\gamma}$ approaches 1 as $\gamma\to \infty$ for essentially all $a,b,c$, this gives us
\begin{equation}
    Q_2^2 = \Lambda^2 \left(\frac{Q_1^2} {\Lambda^2}\right)^{\delaepa^{\beta_0}} \left[\delaepa^{\beta_0}e^{1-\delaepa^{\beta_0}}\right]^{1/\gamma}  \left[\left\{\gamma\ln(Q_1^2/\Lambda^2) - 1\right\}^{1-\delaepa^{\beta_0}}\right]^{1/\gamma}\underset{\gamma\to\infty}{\longrightarrow}
    \Lambda^2 \left(\frac{Q_1^2} {\Lambda^2}\right)^{\delaepa^{\beta_0}} 
\end{equation}
as expected.

\bibliography{bibliography}
\end{document}